\documentclass[fleqn,usenatbib]{mnras}
\usepackage{newtxtext,newtxmath}
\usepackage[T1]{fontenc}
\usepackage{ae,aecompl}

\usepackage{graphicx}	% Including figure files
\usepackage{amsmath}	% Advanced maths commands
\usepackage{amssymb}	% Extra maths symbols

\usepackage{natbib}
\usepackage{color}
\usepackage{rotating}
\usepackage{url}
\usepackage{ifthen}
\usepackage{bm}
\usepackage{bbold}
\usepackage{aas_macros}
\usepackage{verbatim}
\usepackage{algorithm}% http://ctan.org/pkg/algorithms
\usepackage{algpseudocode}% http://ctan.org/pkg/algorithmicx
\usepackage{enumitem}

\newcommand{\mvec}[1]{\bm{#1}}
\newcommand{\mmat}[1]{\mathbf{#1}}

\newcommand{\myvec}[1]{\mvec{#1}}
\newcommand{\mymat}[1]{\mmat{#1}}
\newcommand\numberthis{\addtocounter{equation}{1}\tag{\theequation}}

\SetSymbolFont{symbols}{bold}{OMS}{cmsy}{b}{n}
\DeclareSymbolFont{bmisymbols}{OML}{cmm}{b}{it}

\title[Fast signal reconstruction without preconditioning]{Wiener filter reloaded: fast signal reconstruction without preconditioning}

\author[D. Kodi Ramanah, G. Lavaux, B. D. Wandelt]{Doogesh Kodi Ramanah$^{1,2}$\thanks{ramanah@iap.fr}, Guilhem Lavaux$^{1,2}$\thanks{lavaux@iap.fr}, Benjamin D. Wandelt$^{1,2}$\\
$^{1}$ Sorbonne Universit\'es, UPMC Univ Paris 6 et CNRS, UMR 7095, Institut d'Astrophysique de Paris, 98 bis bd Arago, 75014 Paris, France\\
$^{2}$ Sorbonne Universit\'es, Institut  Lagrange  de  Paris  (ILP),  98  bis bd Arago, 75014 Paris, France\\
}

\date{Accepted XXX. Received YYY; in original form ZZZ}

\pubyear{2017}

\begin{document}
\label{firstpage}
\pagerange{\pageref{firstpage}--\pageref{lastpage}}
\maketitle

\begin{abstract}
We present a high performance solution to the Wiener filtering problem via a formulation that is dual to the recently developed messenger technique. This new dual messenger algorithm, like its predecessor, efficiently calculates the Wiener filter solution of large and complex data sets without preconditioning and can account for inhomogeneous noise distributions and arbitrary mask geometries. We demonstrate the capabilities of this scheme in signal reconstruction by applying it on a simulated cosmic microwave background (CMB) temperature data set. The performance of this new method is compared to that of the standard messenger algorithm and the preconditioned conjugate gradient (PCG) approach, using a series of well-known convergence diagnostics and their processing times, for the particular problem under consideration. This variant of the messenger algorithm matches the performance of the PCG method in terms of the effectiveness of reconstruction of the input angular power spectrum and converges smoothly to the final solution. The dual messenger algorithm outperforms the standard messenger and PCG methods in terms of execution time, as it runs to completion around 2 and 3-4 times faster than the respective methods, for the specific problem considered.
\end{abstract}


%%%%%%%%%%%%%%%%%%%%%%%%%%%%%%%%%%%%%%%%%%%%%%%%%%

%%%%%%%%%%%%%%%%% BODY OF PAPER %%%%%%%%%%%%%%%%%%

\section{Introduction}
\label{section1}

In an era of precision cosmology, the data analysis of state-of-the-art galaxy redshift surveys and CMB experiments with unprecedented levels of sensitivity and resolution poses complex numerical challenges. Fast and robust methods are therefore required to render the data analysis of these large and complex data sets computationally tractable. 

One of the most frequently encountered and ubiquitous problems in astrophysics and cosmology (and many other fields in science) is signal reconstruction from noisy data. Solutions to this problem have therefore been researched extensively for the last two centuries \citep*[e.g.,][]{gauss1809theoria, jaynes1957information, kalman1960new, EW12}. The Wiener filter \citep{wiener1949extrapolation} has emerged as a standard tool for the analysis of large data sets for the inference of high dimensional signals, such as the large-scale structures and CMB problems. It has therefore been employed for large-scale structure analysis problems such as inferring the three dimensional density field from observations \citep*[e.g.,][]{zaroubi1999wiener, zaroubi2002unbiased, erdogdu20042df, erdogdu2006reconstructed, kitaura2008bayesian, kitaura2009cosmic, jasche2010bayesian, jasche2015matrix}. In the analysis of CMB data sets, the Wiener filter has been applied to a range of problems, such as the joint inference of temperature fluctuations and power-spectra \citep*[e.g.,][]{wandelt2004global, eriksen2004power, jewell2004application, odwyer2004bayesian, smith2007background, larson2007estimation, EW12, bunn2016pure}. 

If we assume a linear model where the data $\myvec{d}$ is a combination of the signal $\myvec{s}$ and noise $\myvec{n}$, i.e.,
\begin{equation}
	\myvec{d} = \myvec{s} + \myvec{n}, 
	\label{eq:data_model}
\end{equation} 
the Wiener filter is defined as the solution to the following equation:
\begin{equation}
	(\mymat{S}^{-1} + \mymat{N}^{-1})\myvec{s}_{\text{\tiny {\textup{WF}}}} = \mymat{N}^{-1}\myvec{d},
	\label{eq:wf_equation}
\end{equation} 
where $\mymat{S}$ and $\mymat{N}$ are the signal and noise covariance matrices, respectively, and $\myvec{s}_{\text{\tiny {\textup{WF}}}}$ is the Wiener filter solution. As is evident from Equation (\ref{eq:wf_equation}), the direct numerical implementation of the Wiener filter requires inversion of dense matrices. This task is rendered intractable by the size of modern data sets from state-of-the-art experiments, and therefore represents a computational bottleneck. While it would be extremely convenient if there existed a common basis set, easily accessible by fast transforms, where both $\mymat{S}$ and $\mymat{N}$ are sparse, this is often not the case as for instance, the signal and noise covariances may be sparse in Fourier or pixel space, respectively. Some previous approaches relied on the assumption of a homogeneous and isotropic noise distribution to find approximate solutions to the Wiener filter equation \citep*[e.g.,][]{hirata2004cross, komatsu2005measuring, mangilli2009nongaussianity}. Alternative approaches for the exact solution to the Wiener filter equation involve complex numerical algorithms like Krylov space methods, such as preconditioned conjugate gradient (PCG) techniques \citep*[eg.,][and references therein]{wandelt2004global, eriksen2004power, kitaura2008bayesian}  for matrix inversion and solving high dimensional systems of linear equations. However, this procedure requires intricate and costly numerical schemes such as preconditioning the linear system with suitable matrices and subsequently demands significant investment in software development. Moreover, this approach involves a further inherent stumbling block as finding efficient preconditioners is a complicated task in itself \citep{oh1999efficient} since matrices are often extremely ill-conditioned in typical CMB problems \citep{eriksen2004power}. The inclusion of polarisation data in the analysis further exacerbates this predicament \citep{larson2007estimation}.

Some previous schemes based on the PCG method involved the use of a combination of block and diagonal preconditioners on large and small angular scales, respectively \citep{eriksen2004power}, or were based on a recursive algorithm, where the conjugate gradient solution on a coarse grid is adopted as the preconditioner on a finer grid \citep{smith2007background}. Although both of these approaches yielded satisfactory performance with the analysis of {\it WMAP} data \citep*[e.g.,][]{eriksen2008joint}, they were found to be too computationally intensive for the high resolution and sensitivity analysis of {\it Planck} data. \cite{seljebotn2014multi} recently proposed a multi-level solver for Gaussian constrained realisations of the CMB, which is fast once a suitable preconditioner is found but requires careful tuning and costly precomputations in terms of both computing power and memory requirements. This approach is therefore less attractive when we have to solve many different systems, each requiring a specific preconditioner.

\cite{EW12} presented an alternative approach by devising the messenger algorithm which bypasses the need for a preconditioner, as described in the following section. The evaluation of the Wiener filter from Equation (\ref{eq:wf_equation}) via both the standard and a new formulation that is dual to the messenger algorithm constitutes the crux of this work. 

The paper is organised as follows. In Section \ref{section2}, the underlying principles of the standard messenger method are outlined, followed by a description of the new dual messenger algorithm. We then test our new technique on an artificially generated CMB data set in Section \ref{section3}, and follow up by investigating its performance in terms of convergence, computation time and stability, and draw comparisons to the standard messenger scheme and the popular PCG method in Section \ref{section4}. Finally, in Section \ref{section5}, we summarise the main aspects of our findings and discuss the areas of applications where the potential of our new algorithm can be fully exploited. In Appendices \ref{cooling_scheme_appendix} and \ref{truncating_scheme_appendix}, we illustrate the rationale behind the schemes implemented in this work for fast convergence, followed by a brief review of the PCG method in Appendix \ref{pcg_appendix}.

\section{The Messenger algorithms}
\label{section2}

\subsection{The standard messenger algorithm}

\cite{EW12} proposed a high precision, iterative algorithm for the solution to the full Wiener filter equation, while being numerically efficient and straightforward to implement. Conceptually, the key idea is to introduce a stochastic auxiliary field $\myvec{t}$, the so-called ``messenger" field, with covariance $\mymat{T}$, where $\mymat{T}$ is proportional to the identity matrix. Taking advantage of the useful property of the identity matrix being invariant under any orthogonal basis transformation, the messenger field acts as an intermediate in transformations between different preferred orthogonal bases, in which signal and noise covariance matrices are expressed conveniently, i.e., are sparse. As a result, although directly applying combinations like $(\mymat{S} + \mymat{N})^{-1}$ may not be possible, we can always apply expressions like $(\mymat{S} + \mymat{T})^{-1}$ and $(\mymat{N} + \mymat{T})^{-1}$, irrespective of the basis chosen to render $\mymat{S}$ and $\mymat{N}$ sparse. Under such a scheme, the information from the data is transmitted to the signal via the messenger field that can be transformed efficiently from one basis representation to another, thereby obviating the requirement to apply the inverse Wiener covariance matrix to data. 

With the introduction of the messenger field $\myvec{t}$, the modified $\chi^2$, where the posterior probability distribution of $\myvec{s}$ is proportional to $\exp\left({-\chi^2/2}\right)$, is as follows:
\begin{equation}
	{\chi^2_{\tiny {T}}} = (\myvec{d} - \myvec{t})^{\dagger} \bar{\mymat{N}}^{-1}  (\myvec{d} - \myvec{t}) + (\myvec{t} - \myvec{s})^{\dagger} \mymat{T}^{-1}(\myvec{t} - \myvec{s}) + \myvec{s}^{\dagger}\mymat{S}^{-1}\myvec{s},
	\label{eq:chi2_messenger}
\end{equation}
where we defined $\bar{\mymat{N}} \equiv \mymat{N} - \mymat{T}$, and we choose the covariance matrix of the auxiliary field $\myvec{t}$ according to $\mymat{T} = \alpha \mathbb{1}$, where $\alpha \equiv \textup{min}(\textup{diag}(\mymat{N}))$. Minimising with respect to $\myvec{s}$ and $\myvec{t}$ leads to the following two equations:
\begin{align}
	\left[\bar{\mymat{N}}^{-1} + (\lambda \mymat{T})^{-1}\right]\myvec{t} &= \bar{\mymat{N}}^{-1}\myvec{d} + (\lambda \mymat{T})^{-1}\myvec{s} \label{eq:messenger_1st_equation}\\
	%\label{eq:messenger_1st_equation}
	\left[\mymat{S}^{-1} + (\lambda \mymat{T})^{-1}\right]\myvec{s} &= (\lambda \mymat{T})^{-1}\myvec{t},
	\label{eq:messenger_2nd_equation}
\end{align}
where we also introduced a scalar parameter $\lambda$ whose purpose is to accelerate convergence. In the limit of $\lambda=1$, the above system of Equations (\ref{eq:messenger_1st_equation}) and (\ref{eq:messenger_2nd_equation}) reduces to the usual Wiener filter Equation (\ref{eq:wf_equation}), as shown in Appendix \ref{cooling_scheme_appendix}. The messenger algorithm basically involves the following steps, as outlined in Algorithm \ref{alg:messenger}. We initialise the vectors $\myvec{s}$ and $\myvec{t}$ with zeros, and choose an initial high value of $\lambda$. We first solve Equation (\ref{eq:messenger_1st_equation}) for the messenger field $\myvec{t}$ in the basis defined by $\mymat{N}$, the noise covariance matrix. Then, we change to a basis where the signal covariance matrix $\mymat{S}$ has a sparse representation, such as Fourier space. Next, we solve for $\myvec{s}$ using Equation (\ref{eq:messenger_2nd_equation}) and the resulting $\myvec{t}$ from the previous step. Finally, we transform the result back to the original basis. The signal reconstruction converges to the Wiener filter solution, i.e., $\myvec{s} \rightarrow \myvec{s}_{\text{\tiny {\textup{WF}}}}$, as $\lambda \rightarrow 1$.

\begin{algorithm}
\begin{algorithmic}[1]
  \Procedure{Messenger}{$\myvec{d}$, $\mymat{N}$, $\mymat{S}, N, L$}
  \State $\myvec{s}_0 = \mathrm{zeros}(N,N)$ \Comment{Initialise $\myvec{s}$ with zeros}
  \State $\myvec{t}_0 = \myvec{d}$ \Comment{Initialise $\myvec{t}$ via an initial guess}
  \State \Comment{Compute the covariance of messenger field $\myvec{t}$}
  \State $\alpha = \mathrm{min}(\mathrm{diag}(\mymat{N}))$ \Comment{such that $\mymat{T} = \alpha \mathbb{1}$} 
  %\State $\mymat{T} = \mathrm{min}(\mathrm{diag}(\mymat{N})) \mathbb{1}$ 
  \State $\bar{\mymat{N}} = \mymat{N} - \mymat{T}$ \Comment{Compute the covariance $\bar{\mymat{N}}$}
  \While{$\lambda=10^4$ $\rightarrow$ $1$}
  	\Repeat 
  	    \State \Comment{Transform to Fourier space, $\mathcal{F}$} 
  	    \State \Comment{$\hat{\myvec{s}}_{\vec{\ell}} = \mathcal{F}\myvec{s}_{\vec{x}} = \left( \frac{L}{N} \right)^2 \sum\limits_{\vec{x}} \omega^{-\vec{x}\cdot\vec{k}} \myvec{s}_{\vec{x}} $}
  	    \State \Comment{where $\omega = \exp({\frac{i2\pi}{N}})$}
        \State $\hat{\myvec{s}}_{i+1,\vec{\ell}} = \left[\mymat{S}^{-1} + \sigma(\lambda \mymat{T})^{-1}\right]_{\vec{\ell}}^{-1}$ 
        \State $\; \; \; \; \; \; \; \; \; \; \; \; \; \; \; \; \; \; \; \; \; \; \; \; \; \; \cdot \mathcal{F}\left[ (\lambda \mymat{T})^{-1}_{\vec{x}} \myvec{t}_{i,\vec{x}} \right]_{\vec{\ell}}$
        \State \Comment{$\sigma = \frac{N^2}{L^4}$ is a numerical factor due to $\mathcal{F}$}
        \State \Comment{$N =$ no. of pixels, $L =$ angular extent $\, \,$}  
        \State \Comment{Transform to pixel space}
       	\State $\myvec{s}_{i+1,\vec{x}} = \mathcal{F}^{-1}(\hat{\myvec{s}}_{i+1,\vec{\ell}})$ 
       	\State $\myvec{t}_{i+1,\vec{x}} = \left[\bar{\mymat{N}}^{-1} + (\lambda \mymat{T})^{-1}\right]_{\vec{x}}^{-1}$
       	\State $\; \; \; \; \; \; \; \; \; \; \; \; \; \; \; \; \; \;  \cdot \left[ \bar{\mymat{N}}^{-1}\myvec{d} + (\lambda \mymat{T})^{-1}\myvec{s}_{i+1} \right]_{\vec{x}}$
       	\State $i \gets i+1$
    \Until{$\left\| \myvec{s}_{i} - \myvec{s}_{i-1} \right\| / \left\| \myvec{s}_{i} \right\| < \epsilon$}
    \State $\lambda \gets \lambda \times \eta$ \Comment{Cooling scheme for $\lambda$}
  \EndWhile 
      \State {$\myvec{s} \rightarrow \myvec{s}_{\text{\tiny {\textup{WF}}}}$} \Comment{as $\lambda = 1$}
    \State \Return{$\myvec{s}_{\text{\tiny {\textup{WF}}}}$}
  \EndProcedure
\end{algorithmic}
\caption{\label{alg:messenger} Messenger algorithm}
\end{algorithm}

If $\epsilon_i$ is the residual at the $i$\textsuperscript{th} step, then at the following $(i+1)$\textsuperscript{th} step, the corresponding residual is
\begin{align}
	\epsilon_{i+1} &= \left[\mymat{S}(\mymat{S} + \lambda \mymat{T})^{-1}\right]\left[\bar{\mymat{N}}(\bar{\mymat{N}}+ \lambda \mymat{T})^{-1}\right]\epsilon_i \label{eq:residual_messenger1} \\
	&= \left[\mymat{S}(\mymat{S} + \lambda \mymat{T})^{-1}\right]\left\lbrace(\mymat{N} - \mymat{T}) \left[ \mymat{N} + (\lambda-1)\mymat{T}\right]^{-1} \right\rbrace\epsilon_i.
	\label{eq:residual_messenger2}
\end{align}
For the case of homogeneous noise, i.e., $\mymat{N} \propto \mathbb{1}$, the system is solved exactly in a single step. A key observation is that $\left|\epsilon_{i+1}\right| < \left|\epsilon_i\right|$ for all $i$ since the terms in brackets are less than unity, resulting in unconditional convergence of the signal reconstruction $\myvec{s}$ to the Wiener filter solution $\myvec{s}_{\text{\tiny {\textup{WF}}}}$. A careful inspection of Equation (\ref{eq:residual_messenger2}) yields the following insight: Convergence is fast for low noise pixels and modes with low signal prior variance while the converse is also true, i.e., the system converges slowly for high prior variance and high noise pixels. In the latter regimes, a high value of $\lambda$ would speed up convergence. \cite{EW12} have shown that it is possible to find a cooling scheme for $\lambda$, where $\lambda \gg 1$ initially, to smoothly bring the algorithm to the final solution ($\lambda = 1$). Here, we reduce $\lambda$ by a constant factor $(1/\eta)$, where $1/2 \leq \eta \leq 1$. The rationale behind the cooling scheme adopted in this work is laid out in Appendix \ref{cooling_scheme_appendix}. 

The standard messenger method has enjoyed considerable success, thereby establishing its credentials as a reliable method for fast Wiener filtering, without having recourse to a preconditioner. Essentially, it speeds up the computation of the Wiener filter when the noise is strongly inhomogeneous and/or the data is masked. \cite{elsner2012fast, EW12} applied the messenger algorithm on CMB data from {\it WMAP} satellite and found the final map to be accurate to about 1 part in $10^5$, compared to standard conjugate gradient solvers. Even with the inclusion of polarisation data in the analysis, the messenger algorithm maintained its efficiency. \cite{mangilli2013optimal} implemented the messenger algorithm to produce Wiener-filtered simulations of non-Gaussian CMB maps with the lensing-integrated Sachs Wolf bispectrum signal. The messenger method was also employed in {\it Planck} data analysis, specifically for inverse covariance filtering of CMB maps at high angular resolutions \citep{23planck2013, 24planck2013}. \cite{jasche2015matrix} implemented a Gibbs sampling adaptation of the messenger algorithm in a simple, easy to implement but efficient algorithm for Bayesian large-scale structure inference, specifically aiming at the joint inference of cosmological density fields and power spectra for linear data models. \cite{anderes2015bayesian} adopted a similar approach in their Bayesian hierarchical modelling of the CMB gravitational lensing. \cite{alsing2016hierarchical} also implemented the Gibbs-messenger sampling adaptation developed by \cite{jasche2015matrix} for Bayesian hierarchical modelling of cosmic shear power spectrum inference and eventually for cosmological parameter inference \citep{alsing2016cosmological}. The works of \cite{jasche2015matrix}, \cite{anderes2015bayesian} and \cite{alsing2016hierarchical} show that the messenger algorithm can be successfully adapted for high resolution conditional Gaussian sampling using Markov Chain Monte Carlo techniques, demonstrating the flexibility of the method.

\subsection{The dual messenger algorithm}

In the above messenger algorithm, we made use of a relation to trivially split the noise into two components: one with a trivial covariance matrix and the other as a fluctuating component over the sky. An alternative approach is to introduce the auxiliary field at the level of the signal in a complementary formalism to the standard messenger framework. Due to the two schemes being complementary to each other, the new algorithm is referred to as the dual messenger algorithm. The corresponding log-posterior to be optimised then becomes

\begin{equation}
	{\chi^2_{\tiny {U}}} = (\myvec{d} - \myvec{s})^{\dagger}\mymat{N}^{-1}  (\myvec{d} - \myvec{s}) + (\myvec{s} - \myvec{u})^{\dagger} \mymat{U}^{-1}(\myvec{s} - \myvec{u}) + \myvec{u}^{\dagger}\bar{\mymat{S}}^{-1}\myvec{u},
	\label{eq:chi2_dual_messenger}
\end{equation}
where, analogous to the standard approach, $\mymat{U} = \nu\mathbb{1}$ with $\nu \equiv \mathrm{min}(\textup{diag}(\mymat{S}))$, and the covariance of the auxiliary field, $\bar{\mymat{S}} \equiv \mymat{S} - \mymat{U}$. We derive the corresponding equations that must be satisfied by $\myvec{s}$ and $\myvec{u}$ at the minimum of ${\chi^2_{\tiny {U}}}$:
\begin{align}
	\left(\mymat{N}^{-1} + \mymat{U}^{-1}\right)\myvec{s} &= \mymat{N}^{-1}\myvec{d} + \mymat{U}^{-1}\myvec{u} \label{eq:dual_messenger_1st_equation}\\
	\left(\mymat{U}^{-1} + \bar{\mymat{S}}^{-1} \right)\myvec{u} &= \mymat{U}^{-1}\myvec{s}.
	\label{eq:dual_messenger_2nd_equation}
\end{align}
The dual messenger algorithm has interesting convergence properties. The amount of reduction of the residual at each iteration is given by
\begin{align}
	\epsilon_{i+1} &= [\mymat{N} (\mymat{N} + \mymat{U})^{-1}] [\bar{\mymat{S}} (\bar{\mymat{S}} + \mymat{U})^{-1}] \epsilon_i \label{eq:residual_dual_messenger1} \\ 
	&= [\mymat{N} (\mymat{N} + \mymat{U})^{-1}] [(\mymat{S} - \mymat{U})\mymat{S}^{-1}] \epsilon_i. 
	\label{eq:residual_dual_messenger2}
\end{align}
This provides the basis for the following mechanism: We artificially truncate the spectrum $\mymat{S}$ to some lower initial value of $\ell_\mathrm{iter}$ that corresponds to a covariance $\mu$ and bring $\ell_\mathrm{iter}$ slowly to $\ell_\mathrm{max}$ corresponding to our final covariance $\nu$. So, essentially, we vary the covariance $\mymat{U}$ via a cooling scheme to bring $\mu \rightarrow \nu$, where, in the limit $\mu = \nu$, the above system of Equations (\ref{eq:dual_messenger_1st_equation}) and (\ref{eq:dual_messenger_2nd_equation}) reduces to the usual Wiener filter Equation (\ref{eq:wf_equation}), as shown in Appendix \ref{truncating_scheme_appendix}.  This results in a redefinition of $\bar{\mymat{S}}$ using the Heaviside function as $\bar{\mymat{S}} = \Theta (\mymat{S} - \mymat{U})$, as described quantitatively in Appendix \ref{truncating_scheme_appendix} (cf. Equations (\ref{eq:heaviside1_appendix}) and (\ref{eq:heaviside2_appendix})). For $\ell \la \ell_\mathrm{iter}$, the ratio given by Equation (\ref{eq:residual_dual_messenger2}) is always convergent. The algorithm consists of the following steps. We initialise the vectors $\myvec{s}$ and $\myvec{u}$ with zeros. We begin iterations with an initial value of $\ell_\mathrm{iter}$ and correspondingly $\mu$, and iterate until $\mu \rightarrow \nu$ at $\ell_\mathrm{iter} = \ell_\mathrm{max}$, i.e., $\myvec{s} \rightarrow \myvec{s}_{\text{\tiny {\textup{WF}}}}$, in accordance with a chosen convergence criterion for each value of $\mu$. In analogy with the messenger technique, we need to adopt a cooling scheme for $\mu$.  The idea is to reduce $\mu$ by a constant factor $(1/\beta)$, where $0< \beta < 1$, which matches the convergence speed of one iteration with the modification in $\mu$. The rationale behind the cooling scheme tailored for the dual messenger algorithm is illustrated in Appendix \ref{truncating_scheme_appendix}.

%\subsection{The hybrid dual messenger algorithm}

However, numerically, the above algorithm does not result in the correct final solution due to the continuous mode of the signal, i.e., the zero eigenvalue in the signal covariance $\mymat{S}$. We therefore require $\mu \rightarrow \nu = 0$ to obtain the proper solution at the end, which cannot be accommodated by the dual messenger scheme above. To remedy this numerical predicament, we introduce an extra degree of freedom, $\alpha$, in the system, where $\alpha \equiv \textup{min}(\textup{diag}(\mymat{N}))$, thereby incorporating aspects of the standard messenger method into the dual messenger scheme. This leads to the following $\chi^2$:
\begin{equation}
	{\chi^2_{\tiny {\xi}}} = (\myvec{d} - \myvec{t})^{\dagger}\bar{\mymat{N}}^{-1}  (\myvec{d} - \myvec{t}) + (\myvec{t} - \myvec{s})^{\dagger} \bm{\xi}^{-1}(\myvec{t} - \myvec{s}) + \myvec{s}^{\dagger}\bar{\mymat{S}}^{-1}\myvec{s},
	\label{eq:chi2_hybrid_messenger}
\end{equation}
where $\bm{\xi} = (\alpha + \mu) \mathbb{1} = \xi \mathbb{1} = \mymat{T} + \mymat{U}$, with $\mu$, $\mymat{T}$, $\bar{\mymat{S}}$ and $\bar{\mymat{N}}$ inheriting their previous definitions. The corresponding set of equations to be solved iteratively is then:
\begin{align}
	\left(\bar{\mymat{N}}^{-1} + \bm{\xi}^{-1}\right)\myvec{t} &= \bar{\mymat{N}}^{-1}\myvec{d} + \bm{\xi}^{-1}\myvec{s} \label{eq:hybrid_messenger_1st_equation}\\
	\left(\bm{\xi}^{-1} + \bar{\mymat{S}}^{-1} \right)\myvec{s} &= \bm{\xi}^{-1}\myvec{t}.
	\label{eq:hybrid_messenger_2nd_equation}
\end{align}
If $\alpha = 0$, we recover the usual dual messenger scheme, while setting $\mu = 0$ yields the standard messenger algorithm. The definition of $\bm{\xi}$ implies that the cooling scheme described above still applies to this hybrid method. As outlined in Algorithm \ref{alg:dual_messenger}, we proceed in similar steps as described above for the previous scheme, except that here we reduce the norm of $\bm{\xi}$ by the factor ($1/\beta$) and iterate until $\xi \rightarrow \alpha$, at which point $\mu = 0$ and we obtain the proper solution as desired.

%\subsubsection*{Hybrid Messenger}

\begin{algorithm}
\begin{algorithmic}[1]
  \Procedure{Dual Messenger}{$\myvec{d}$, $\mymat{N}$, $\mymat{S}, N, L$}
  \State $\myvec{s}_0 = \mathrm{zeros}(N,N)$ \Comment{Initialise $\myvec{s}$ with zeros}
  \State $\myvec{t}_0 = \myvec{d}$ \Comment{Initialise $\myvec{t}$ via an initial guess $\,$}
  \State \Comment{Compute the covariance of auxiliary field $\myvec{t}$}
  %\State $\mymat{T} = \mathrm{min}(\mathrm{diag}(\mymat{N})) \mathbb{1}$ 
  \State $\alpha = \mathrm{min}(\mathrm{diag}(\mymat{N}))$ \Comment{such that $\mymat{T} = \alpha \mathbb{1}$} 
  \State $\bar{\mymat{N}} = \mymat{N} - \mymat{T}$ \Comment{Compute the covariance $\bar{\mymat{N}}$}  
  %\State \Comment{Compute the final truncation $\nu$}
  %\State $\nu = \mathrm{min}(\mathrm{diag}(\mymat{S}))$ 
  %\While{$\mu > \nu$}
  \While{$\xi = (\alpha + \mu) \rightarrow \alpha$}  
  	\State $\mymat{U} = (\sigma \mu) \mathbb{1}$ \Comment{Compute covariance $\mymat{U}$} 
  	\State \Comment{As in Algorithm \ref{alg:messenger}, factor of $\sigma$ due to $\mathcal{F}$}
  	\State $\bar{\mymat{S}} = \Theta (\mymat{S} - \mymat{U})$ \Comment{Compute covariance $\bar{\mymat{S}} \:$} 
  	%\State $\nu \gets \sigma \mu \mathbb{1}$
  	\Repeat
  	    %\State \Comment{$\mathcal{F} =$ Fourier transform}
  	    \State \Comment{Moving to Fourier space, $\mathcal{F}$}
  	    %\State $\hat{\myvec{s}}_{i+1} = \mathcal{F}(\myvec{s}_i)$
  	    \State $\hat{\myvec{t}}_{i+1,\vec{\ell}} = \left[ (\bar{\mymat{S}}^{-1} + \sigma \bm{\xi}^{-1})^{-1} \sigma \bm{\xi}^{-1} \right]_{\vec{\ell}} \mathcal{F}(\myvec{t}_{i,\vec{x}}) $
        \State \Comment{Transform to pixel space $\, \, \, \, \,$}
       	\State $\myvec{s}_{i+1,\vec{x}} = \mathcal{F}^{-1}(\hat{\myvec{t}}_{i+1,\vec{\ell}})$ 
       	\State $\myvec{t}_{i+1,\vec{x}} = \left(\mymat{N}^{-1} + \bm{\xi}^{-1}\right)_{\vec{x}}^{-1}$
       	\State $\; \; \; \; \; \; \; \; \; \; \; \; \; \; \; \; \; \; \cdot \left( \mymat{N}^{-1}\myvec{d} + \bm{\xi}^{-1}\myvec{s}_{i+1} \right)_{\vec{x}}$
       	\State $i \gets i+1$
    \Until{$\left\| \myvec{s}_{i} - \myvec{s}_{i-1} \right\| / \left\| \myvec{s}_{i} \right\| < \epsilon$}
    \State $\xi \gets \xi \times \beta$ \Comment{Cooling scheme for $\xi$}
    \State $\mu \gets (\xi - \alpha)/\sigma$ \Comment{Compute resulting $\mu$}
  \EndWhile 
      \State {$\myvec{s} \rightarrow \myvec{s}_{\text{\tiny {\textup{WF}}}}$} \Comment{as $\xi = \alpha$, $\mu = 0$}
    \State \Return{$\myvec{s}_{\text{\tiny {\textup{WF}}}}$}
  \EndProcedure
\end{algorithmic}
\caption{\label{alg:dual_messenger} Dual messenger algorithm}
\end{algorithm}

\section{Application to Cosmic Microwave background}
\label{section3}

\begin{figure*}
	\centering
		{\includegraphics[width=\hsize,clip=true]{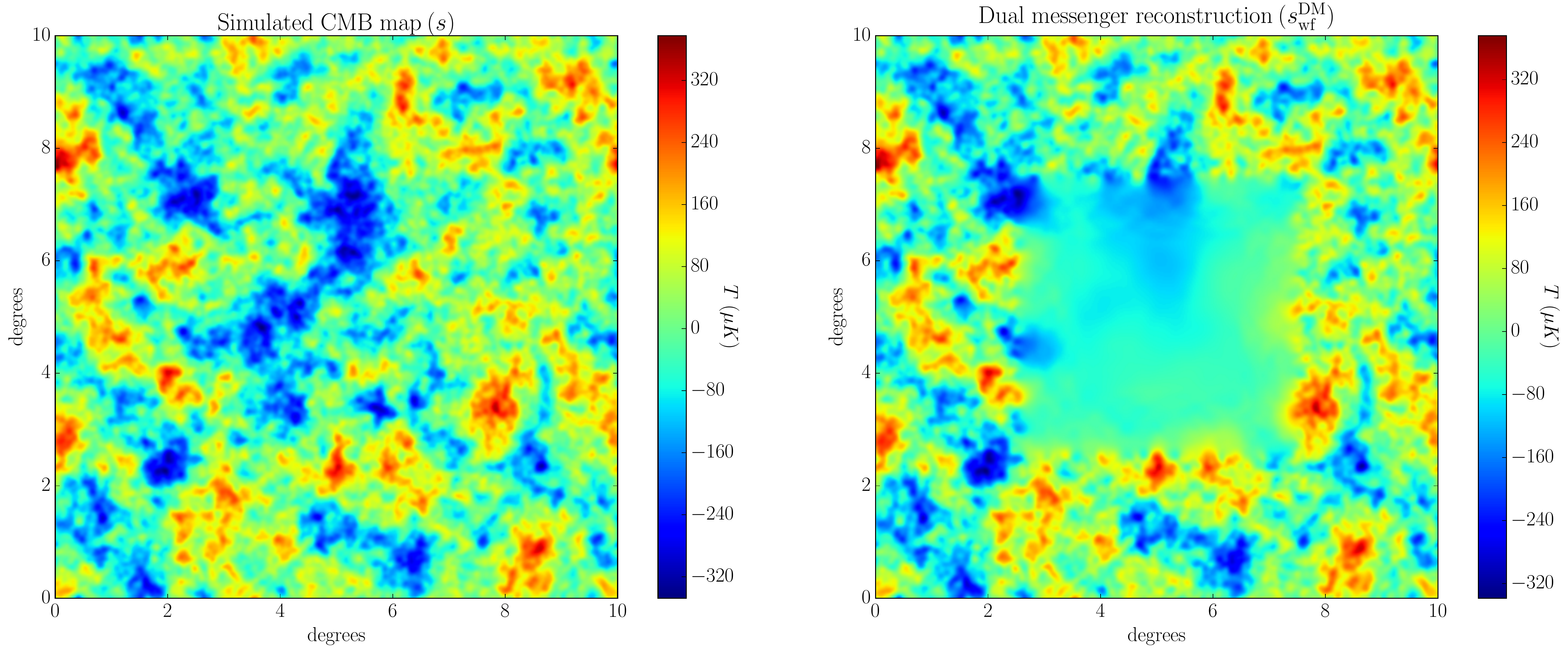}}
	\caption{The simulated CMB map and the reconstruction via the dual messenger algorithm. The left-hand panel shows the simulated map, which is subsequently contaminated by white noise, with the central square patch masked by extremely high noise covariance. The right-hand panel illustrates the corresponding reconstructed map obtained via the dual messenger technique. The messenger and PCG reconstructions are not shown as they look similar to the dual messenger reconstructed map. The residual maps displayed in Figure \ref{fig:residual_map}, however, help to discern the differences between the various reconstructions.}
	\label{fig:recon_map}
\end{figure*}

\begin{figure}
	\centering
		{\includegraphics[width=\hsize,clip=true]{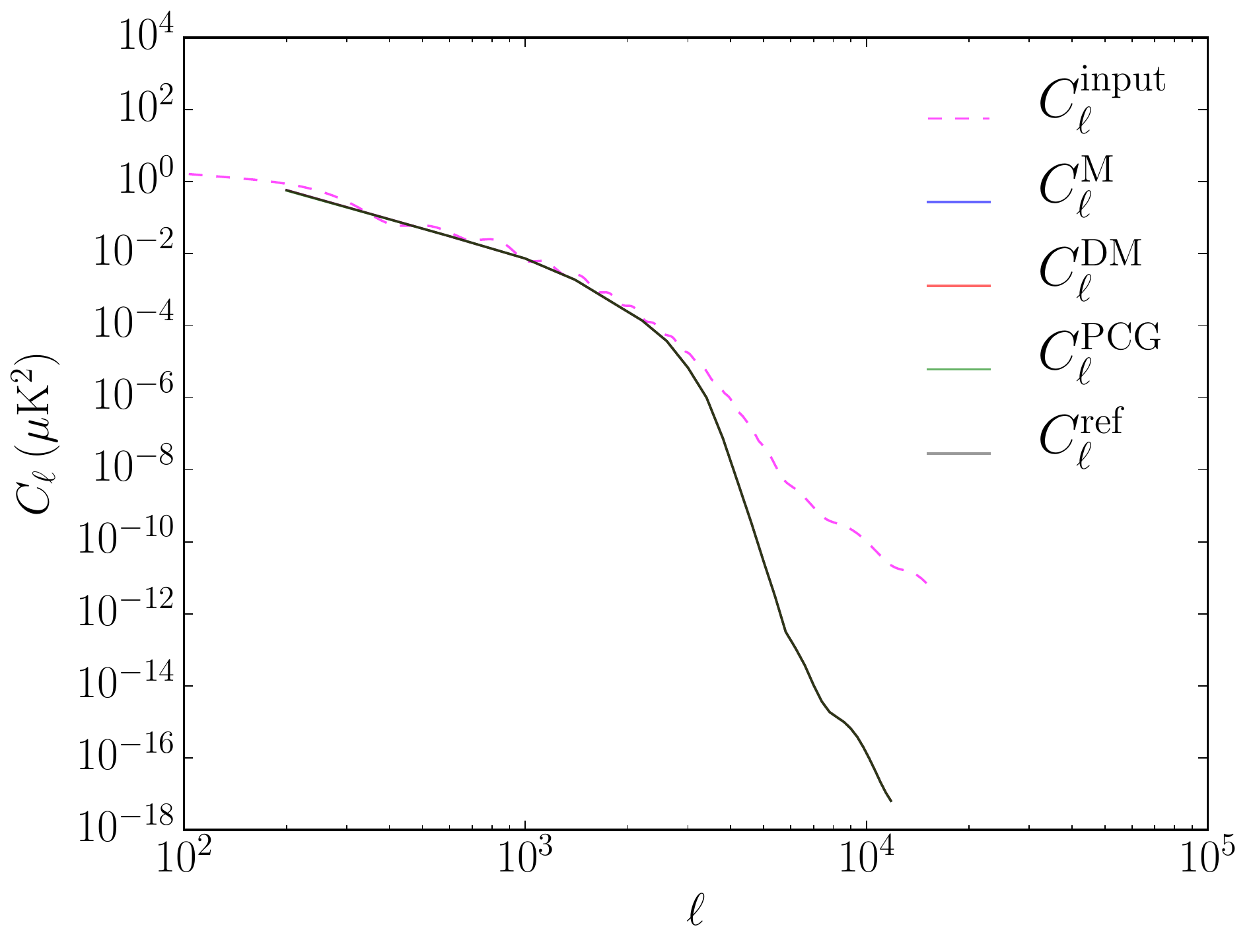}}
	\caption{Reconstructed power spectra computed using the different algorithms. The dashed line indicates the input angular power spectrum, $C_\ell^\mathrm{input}$, from which the CMB signals are drawn. The power spectra recovered by the three algorithms are all in good agreement with the reference power spectrum computed using PCG method with $\epsilon = 10^{-9}$ on all scales. The deviations on the small scales from the input power spectrum are due to the characteristic feature of a Wiener filtered signal, where the power on small scales, in the low signal to noise regime, is suppressed.}
	\label{fig:C_l_recon}
\end{figure}

\begin{figure*}
	\centering
		{\includegraphics[width=\hsize,clip=true]{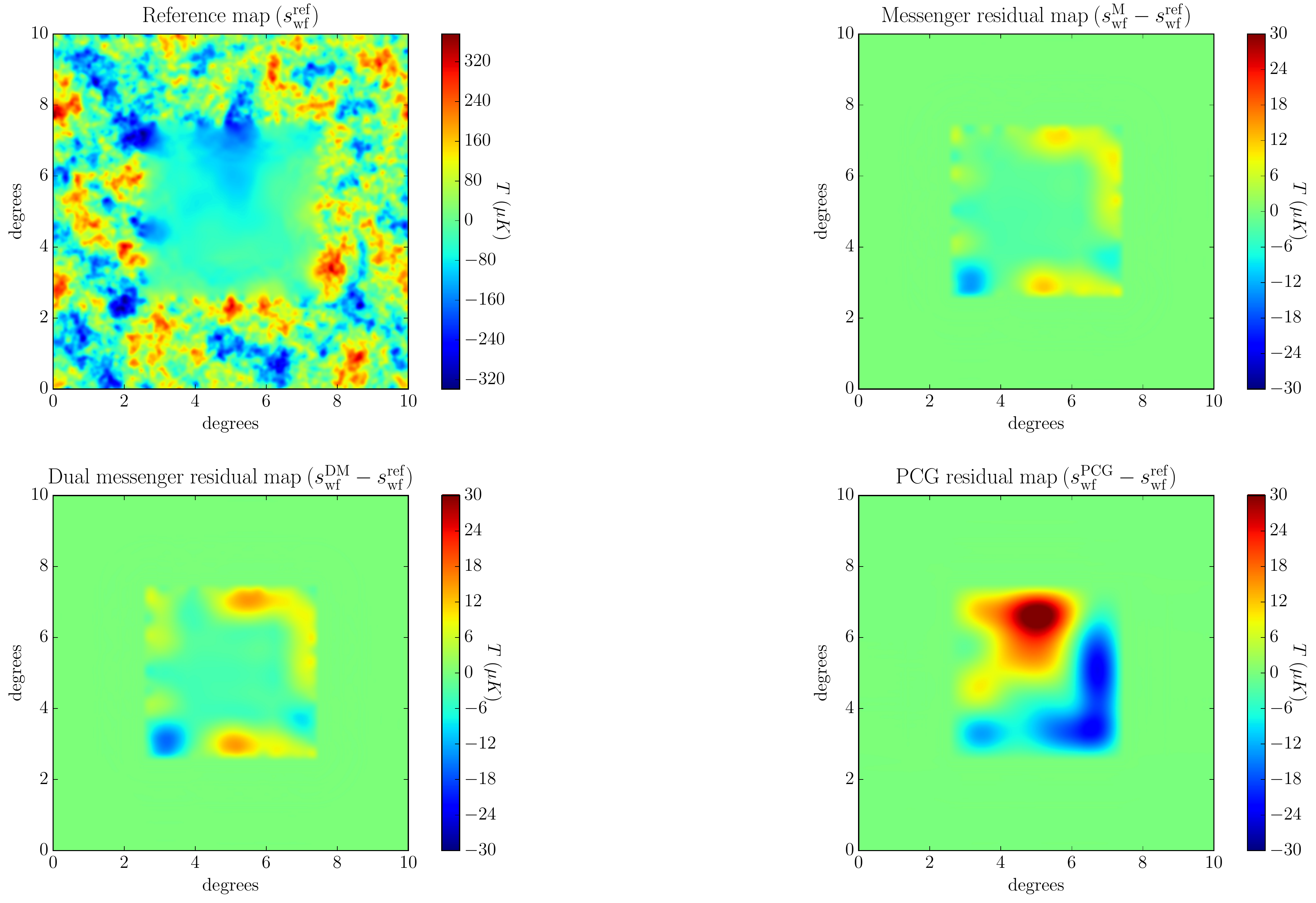}}
	\caption{The reference map and residual maps from the three algorithms. The top left-hand panel depicts the reference map computed using the PCG scheme with $\epsilon=10^{-9}$, while the other panels illustrate the residual maps yielded by the three methods, generated by computing the difference between the reference map and the corresponding reconstructed maps over the full sky. The messenger approach produces the least amount of residuals, around $3\%$, while the dual messenger and PCG schemes result in approximately $4\%$ and $9\%$ residuals, respectively. For the messenger reconstructions, the residuals lie mostly on the edges of the mask.}
	\label{fig:residual_map}
\end{figure*}

\begin{figure}
	\centering
		%{\includegraphics[width=\hsize,clip=true]{cauchy_high_noise_mod3.pdf}}
		{\includegraphics[width=\hsize,clip=true]{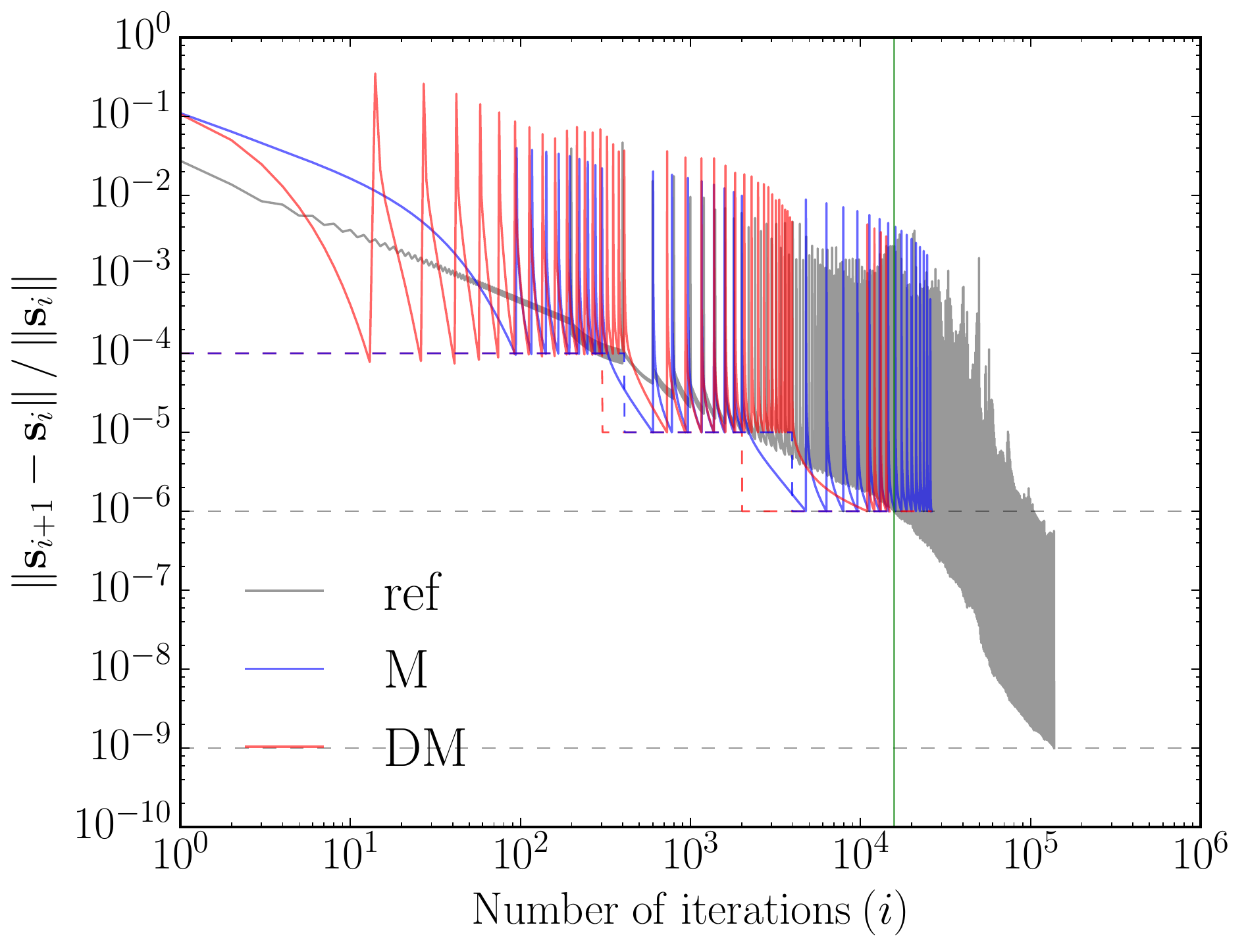}}
	\caption{Variation of the residual error, given by the Cauchy criterion, with number of iterations for the different methods. The Cauchy convergence criteria $\epsilon_i$ imposed for the different regimes of the cooling schemes for $\lambda$ and $\mu$ are also displayed, along with the corresponding thresholds $\epsilon$ used for the two PCG methods, in dashed lines. The vertical green line denotes the convergence point of the PCG scheme, with its convergence behaviour already represented by the reference PCG method. The dual messenger algorithm requires the smallest number of iterations to converge to the final solution than the other two methods.}
	\label{fig:cauchy}
\end{figure}

\begin{figure}
	\centering
		{\includegraphics[width=\hsize,clip=true]{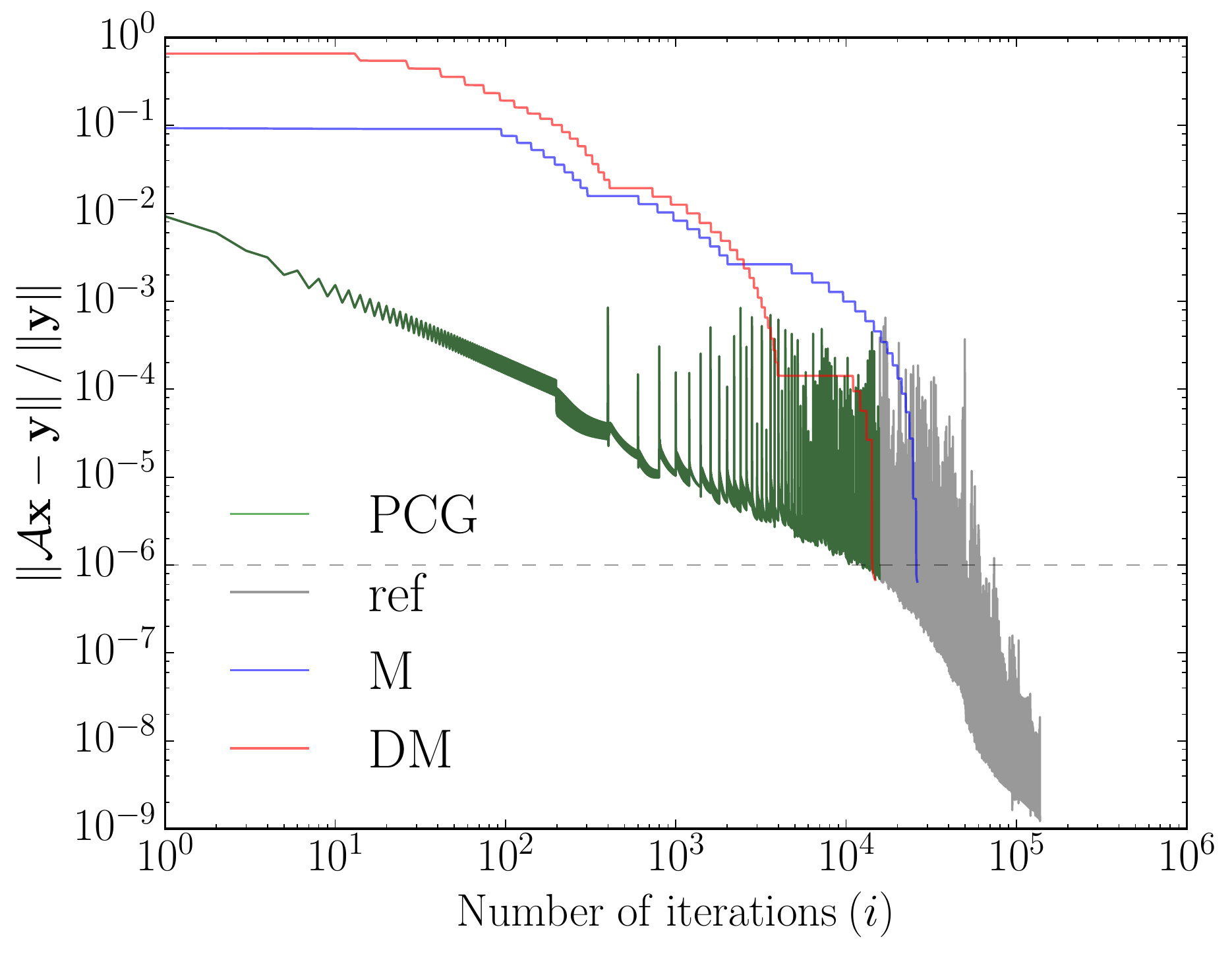}}
	\caption{Variation of the residual error, given by $\left \| \mymat{\mathcal{A}}\myvec{x} - \myvec{y} \right \| / \left \| \myvec{y} \right \|$, with number of iterations for the different algorithms. The corresponding residual errors for all methods drop below the convergence thresholds adopted, demonstrating the consistency of our computations dictated by the Cauchy criteria (cf. Figure \ref{fig:cauchy}).}
	\label{fig:pcg_residual}
\end{figure}

\begin{figure}
	\centering
		{\includegraphics[width=\hsize,clip=true]{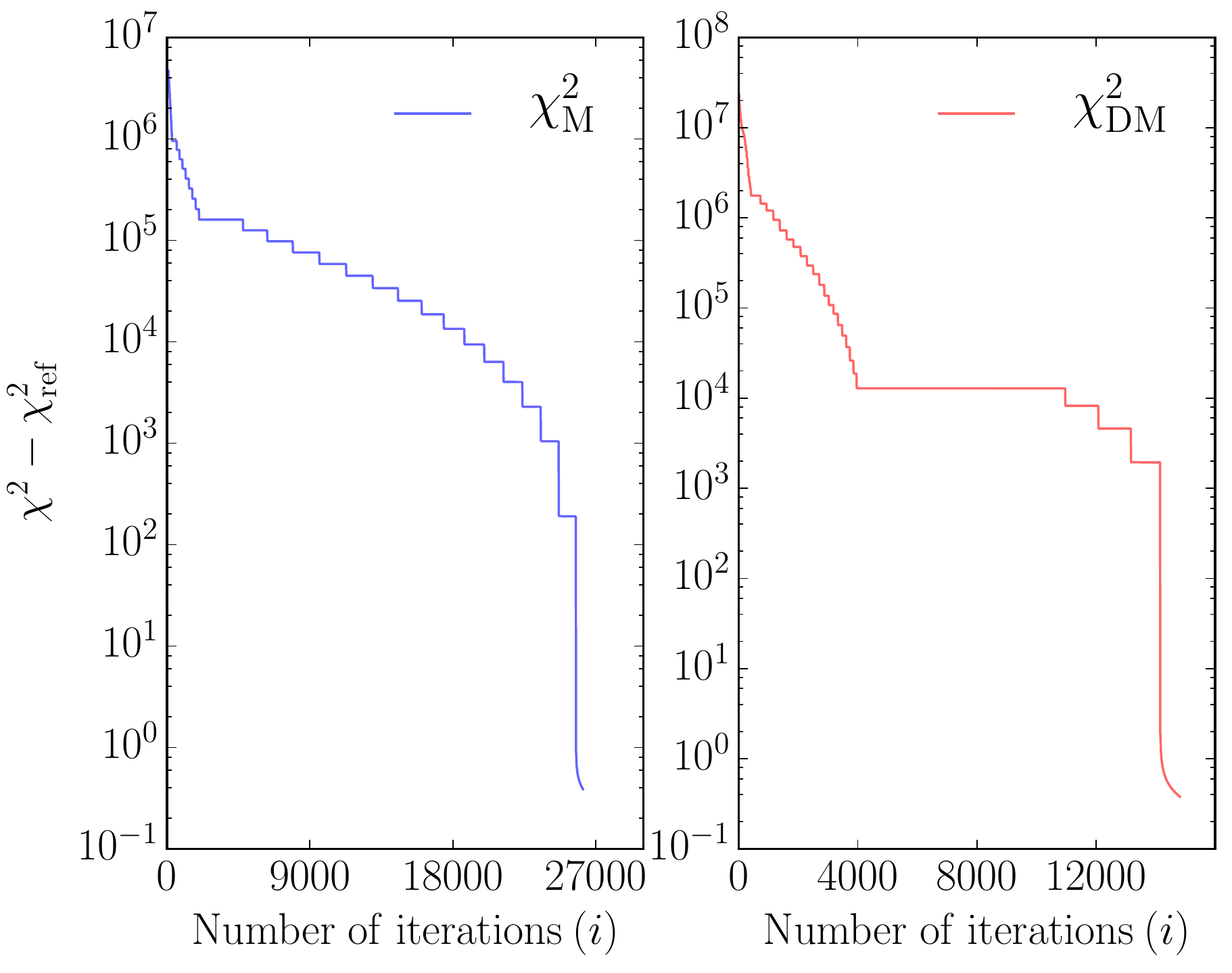}}
	\caption{Variation of $\chi^2$ with number of iterations for the two messenger methods. In the left panel, we find that the $\chi^2_\mathrm{M}$ of the messenger algorithm drops rapidly with the cooling scheme for $\lambda$ adopted, as expected from the earlier discussion. In the right panel, the $\chi^2_\mathrm{DM}$ of its dual counterpart displays a similar behaviour. In both cases, the final solution matches the $\chi^2_\mathrm{ref}$ of the PCG method with $\epsilon = 10^{-9}$.}
	\label{fig:chi_squared}
\end{figure}

We generated artificial CMB data, in a realistic scenario, by drawing Gaussian random fields on a 2d flat sky with grid resolution, $N_{\textup{pixels}} = 512^2$, and angular extent, $L = 10.0$ degrees. This true simulated map is contaminated by Gaussian white noise with covariance matrix, $\mymat{N} = (64.0 \: \mu \mathrm{K^2}) \: \mathbb{1}$, via a linear model as given in Equation (\ref{eq:data_model}). We made use of CAMB\footnote{http://camb.info} \citep{camb1999} to generate the input angular power spectrum from which the CMB signals are drawn. We assume a standard $\Lambda$CDM cosmology with the set of cosmological parameters ($\Omega_m = 0.32$, $\Omega_\Lambda = 0.69$, $\Omega_b = 0.05$, $h = 0.67$, $\sigma_8 = 0.83$, $n_s = 0.97$) from {\it Planck} \citep{13planck2015}. A portion of the flat sky is masked by imposing a central square patch of angular extent $5.0$ degrees with a noise covariance of $10^6 \mymat{N}$. The three different methods, messenger, dual messenger and PCG, are applied to this set-up to investigate their effectiveness and efficiency of signal reconstruction. We implement the same ``weak'' Cauchy convergence criterion, $\left \| \myvec{s}_{i+1} - \myvec{s}_{i} \right\| / \left\| \myvec{s}_{i} \right\| < \epsilon$, where $\epsilon = 10^{-6}$, in all the algorithms, to ensure unbiased results. As a reference, we make use of the PCG method with the more stringent $\epsilon = 10^{-9}$ to provide results against which the other methods can be compared. Currently regarded as the standard Wiener filtering technique by the scientific community, the PCG method is the natural choice for providing a reference solution in the absence of an exact analytic solution. By imposing a more stringent convergence criterion, we ensure that the PCG solution is close to the ground truth.

The true CMB simulated signal and the reconstructed map obtained using the dual messenger algorithm are displayed in Figure \ref{fig:recon_map}. The characteristic feature of Wiener filtering, i.e., the extrapolation of signal into the mask, can be distinctly observed from the reconstructed map. At an initial cursory glance, the reconstructed maps from all three methods appear rather similar, and hence, only the dual messenger reconstruction is displayed in Figure \ref{fig:recon_map}. This observation holds for the low noise regions but under scrutiny, the reconstruction in masked regions shows some slight differences. The residual maps depicted in Figure \ref{fig:residual_map}, generated by computing the difference, over the full sky, between our reference map and the corresponding reconstructed map obtained using each method, help to discern these differences. The messenger technique provides the most accurate reconstruction in the masked region as illustrated by its residual map, with less than $3\%$ residuals. The dual messenger and PCG reconstructions yield around $4\%$ and $9\%$ residuals, respectively, with the residuals obtained via the former algorithm being concentrated on the edges of the mask.

The effectiveness of reconstruction of the dual messenger technique on both small and large scales is also manifest from Figure \ref{fig:C_l_recon} which shows the reconstructed power spectra from the different algorithms. The corresponding power spectra recovered via the messenger, dual messenger and PCG methods are all in good agreement, on all scales, with the reference one, computed using the PCG scheme with the more stringent convergence criterion of $\epsilon = 10^{-9}$. The characteristic feature of Wiener filtering where the recovered power on the small scales in the low signal to noise regime is suppressed is also observed from Figure \ref{fig:C_l_recon}. Standard Wiener filtering algorithms usually encounter difficulties in dealing with masked regions having infinite noise. By masking $25\%$ of the simulated map with noise of the order $10^6$ higher than the unmasked region, we are investigating the worst case performance aspect of the dual messenger algorithm. Hence, we do not expect the algorithm to encounter any difficulties in Wiener filtering maps with high levels of inhomogeneous noise.

The dual messenger algorithm is therefore relevant for current and future high resolution CMB experiments such as South Pole Telescope, Advanced ACTPol, Simons Observatory and CMB-S4. The application of the algorithm can be extended to problems involving the polarisation of the CMB. The formalism remains unchanged for spin field reconstruction, although the numerical implementation is less trivial due to the correlation between the temperature and polarisation components of the signal covariance. The Wiener filtering of polarised CMB data with more complex noise models will be subjected to future investigation to further showcase the efficiency of the dual messenger algorithm in treating complex CMB problems.

\section{Numerical Convergence and Stability}
\label{section4}

We now investigate the convergence properties of the dual messenger algorithm and also provide a more in depth study of the performance of the standard messenger scheme. We quote the usual statistics for convergence and the change in $\chi^2$ of the posterior probability density between successive iterations for each method, so that unbiased comparisons of their efficiency and effectiveness can be drawn. 

Figure \ref{fig:cauchy} shows the residual error at each iteration for the different methods, as a function of the number of iterations. For the messenger algorithm, we relax the convergence criterion $\epsilon$ for high values of $\lambda$ and reduce $\epsilon$ to $10^{-6}$ as $\lambda \rightarrow 1$ in a three-step procedure, as depicted in Figure \ref{fig:cauchy} in dashed red lines. Here, we adopt a cooling scheme where $\lambda$ is reduced by a constant factor of $4/3$, i.e, $\eta = 3/4$, until $\lambda = 1$. This consequently ensures that the $\chi^2$ decreases rapidly, thereby bringing us closer to the final solution with a smaller number of iterations, resulting in faster convergence. 

For the dual messenger algorithm, we implemented a cooling scheme for $\xi$ with $\beta = 3/4$, which results in fast convergence while providing accurate results. So, we pick an initial value of $\mu$ corresponding to an initial value of $\ell_\mathrm{iter}$, and therefore $\xi$, and iterate until convergence, as dictated by a given Cauchy criterion. We repeat this iterative procedure, in accordance with the aforementioned cooling scheme, until $\xi = \alpha$, at which point we have the Wiener filter solution. Again, we impose less stringent convergence criterion at higher values of $\xi$, as shown in Figure \ref{fig:cauchy} in dashed blue lines. 

Figure \ref{fig:cauchy} essentially illustrates the convergence of the different methods. The dual messenger algorithm requires slightly fewer iterations to converge to the final solution than the PCG approach, while the messenger technique requires nearly twice as many iterations as its dual counterpart for convergence. In terms of wall-clock times, the dual messenger has the definite upper hand, as it runs to completion in around 258 seconds whereas the messenger and PCG methods have corresponding wall-clock times of roughly 435 seconds and 663 seconds, i.e., reconstruction via the dual messenger algorithm is nearly three and two times faster than the PCG and messenger methods, respectively. All computations were run on a single core of an Intel Core i5-4690 CPU (3.50 GHz). Both messenger algorithms possess the same algorithmic complexity and memory requirements. They require two Fourier transforms, $\mathcal{O}(N_{\textup{pixels}} \log N_{\textup{pixels}})$, and two scalar multiplications corresponding to algebraic operations of $\mathcal{O}(N_{\textup{pixels}})$, per iteration. In terms of memory requirements, two vectors of size $N_{\textup{pixels}}$ must be temporarily stored in memory. In comparison, the PCG method requires three Fourier transforms and ten scalar multiplications per iteration, and temporary storage of eight vectors of dimension $N_{\textup{pixels}}$ in memory.

We also performed a consistency check by verifying the variation of the residual error given by $\left \| \mymat{\mathcal{A}}\myvec{x} - \myvec{y} \right \| / \left \| \myvec{y} \right \|$. This is usually adopted as a convergence criterion for PCG computations. The variation of this residual error, analogous to Figure \ref{fig:cauchy}, is illustrated in Figure \ref{fig:pcg_residual}. The corresponding residual errors obtained via the different algorithms all drop below their respective Cauchy convergence thresholds implemented. The oscillatory behaviour of the PCG solution, displayed in Figures \ref{fig:cauchy} and \ref{fig:pcg_residual}, is mainly due to the re-initialisation step after every two hundred iterations in the algorithm, as described in Appendix \ref{pcg_appendix}. However, there are also some oscillations in the residual errors due to the PCG method being susceptible to instabilities sourced by numerical noise. The oscillations in the residual errors in Figure \ref{fig:cauchy} of the two messenger solutions are however due to their respective cooling schemes, resulting in transitions in the systems of equations with the varying covariances of the auxiliary fields (cf. Equations (\ref{eq:relative_error_M_appendix}) and (\ref{eq:final_relative_error_DM_appendix})), with the peaks produced coinciding with these transitions. It is important to note that the residual errors always drop sharply after the peaks, thereby demonstrating the unconditional stability of the messenger algorithms. From numerical experiments, the messenger techniques have proven to be far more stable than the PCG method for nearly degenerate systems. 

Due to the Wiener filter being the maximum a posteriori solution, the $\chi^2$ of the intermediate solution can be regarded as a useful convergence diagnostic. The $\chi^2$ variation as a function of number of iterations is displayed in Figure \ref{fig:chi_squared}, with the left and right panels correspondingly showing the convergence of the messenger and dual messenger algorithms. The cooling scheme for $\lambda$ implemented in the standard messenger technique causes the $\chi^2_\mathrm{M}$ to drop rapidly with each change in $\lambda$. Intuitively, this decrease of $\lambda$ via a series of such steps seems reasonable since the $\chi^2$ decreases sharply with each change in $\lambda$, and then reaches a plateau for a given $\lambda$ until the latter decreases further, as evidenced in Figure 2 of \cite{elsner2012fast}. The $\chi^2_\mathrm{DM}$ of the dual messenger algorithm has a similar behaviour. The $\chi^2$ in both cases finally attains the $\chi^2_\mathrm{ref}$ of the reference PCG method with $\epsilon = 10^{-9}$, with $\Delta \chi^2_{\mathrm{M}} / \chi^2_{\mathrm{ref}} = 9.2 \times 10^{-6}$ and $\Delta \chi^2_{\mathrm{DM}} / \chi^2_{\mathrm{ref}} = 9.0 \times 10^{-6}$, where $\chi^2_{\mathrm{ref}} = 4.2 \times 10^4$.

\begin{figure}
	\centering
		{\includegraphics[width=\hsize,clip=true]{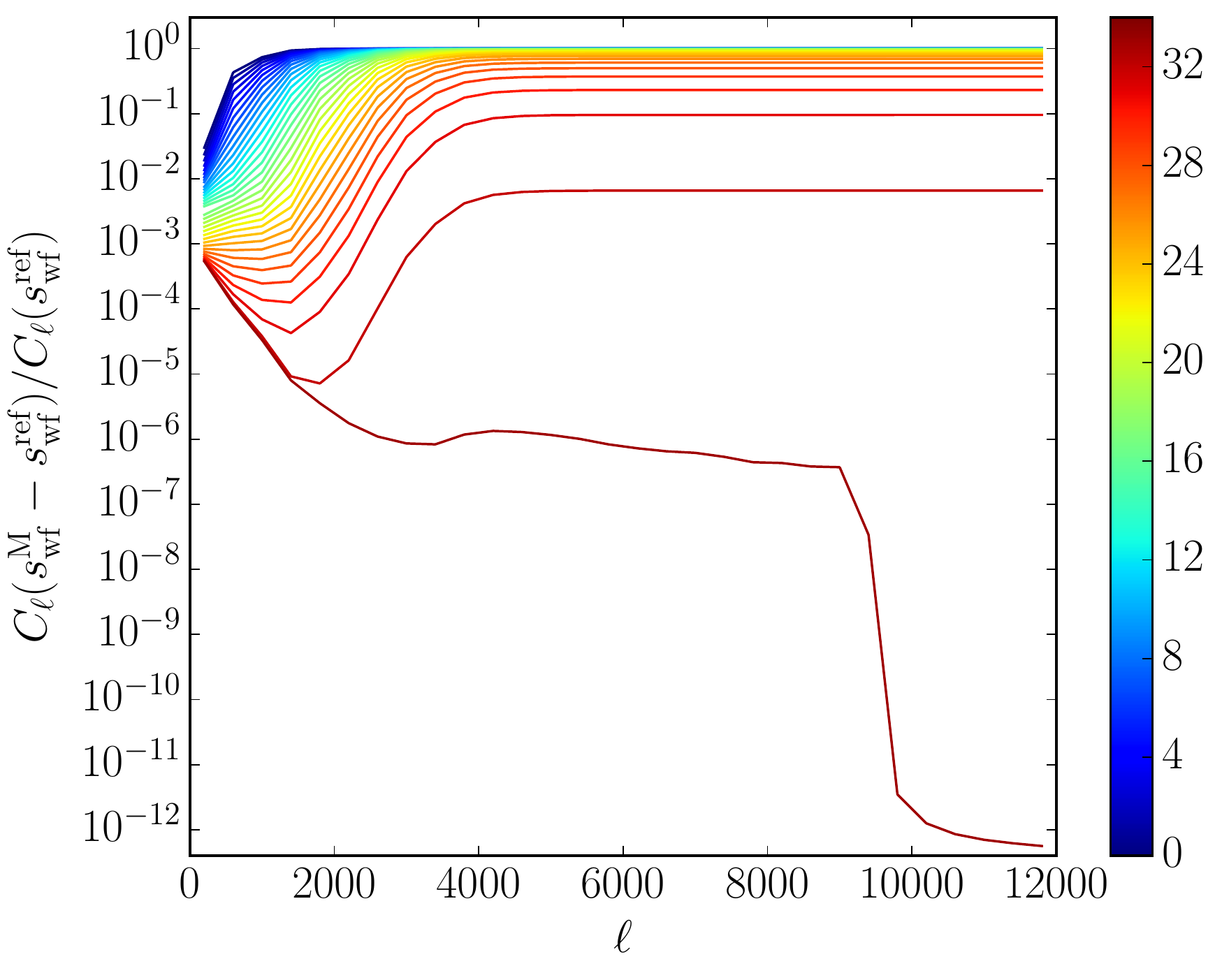}}
	\caption{Convergence rate by frequency bin for the messenger method. This illustrates the relative error as a function of scale, $\ell$. Each line in the figure corresponds to the relative error for a specific value of the scalar parameter $\lambda$. The messenger algorithm converges smoothly and in nearly uniform fashion on small and intermediate scales, with the relative error dropping till below $10^{-6}$. However, for larger scales, the relative error is reduced by lower extent, but stays below $10^{-3}$ for the largest scales. For $\lambda = 1$, the behaviour is similar to that displayed by PCG (cf. Figure \ref{fig:conv_test_PCG}).} %The relative spacing of the lines as $\lambda \rightarrow 1$ implies that convergence is faster as $\lambda$ gets smaller.}
	\label{fig:conv_test_M}
\end{figure}

\begin{figure}
	\centering
		{\includegraphics[width=\hsize,clip=true]{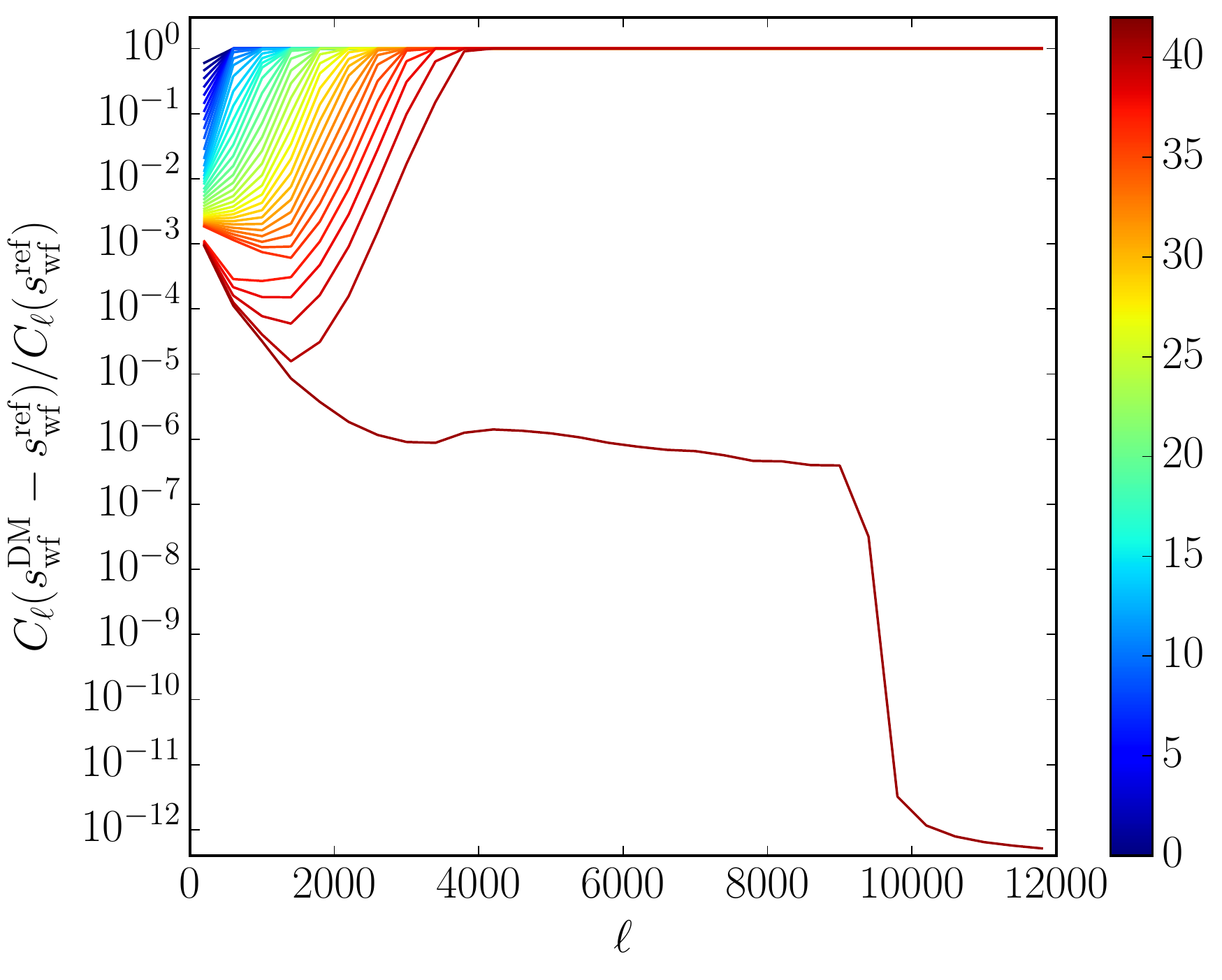}}
	\caption{Convergence rate by frequency bin for the dual messenger method. Same as Figure \ref{fig:conv_test_M}, except that here each line in the figure corresponds to a given value of $\xi$, displaying the hierarchical nature of this scheme. The final relative error is sufficiently low on all scales. A quantitative explanation of the convergence behaviour is given in Appendix \ref{truncating_scheme_appendix}.}
	\label{fig:conv_test_DM}
\end{figure}

\begin{figure}
	\centering
		{\includegraphics[width=\hsize,clip=true]{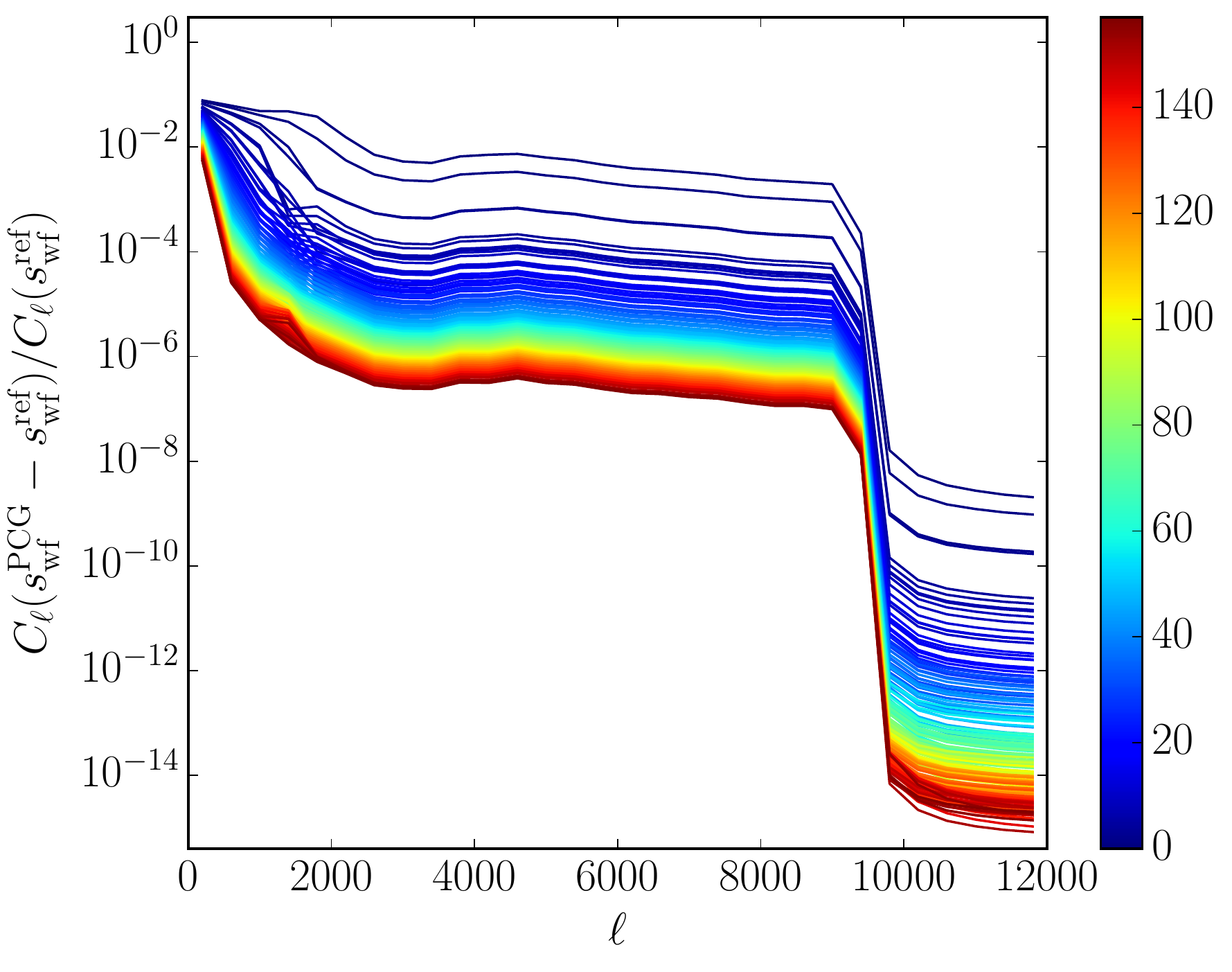}}
	\caption{Convergence rate by frequency bin for the PCG method. Same as Figure \ref{fig:conv_test_M}. The relative error as a function of scale for every 100\textsuperscript{th} iteration is illustrated. The PCG method has a lower relative error than the dual messenger algorithm on the small scales, but the behaviour on the largest scales remains inferior to that of both messenger methods. The relative spacing of the lines indicates that convergence slows down as the PCG method progresses in iterations, and this is in agreement with its variation of the residual error with number of iterations, as displayed by the green line in Figure \ref{fig:cauchy}.}
	\label{fig:conv_test_PCG}
\end{figure}

Another important convergence diagnostic is the variation of the relative error, $C_{\ell}(\myvec{s}_{\text{\tiny {\textup{WF}}}} - \myvec{s}_{\text{\tiny {\textup{WF}}}}^{\textrm{ref}})/ C_{\ell}(\myvec{s}_{\text{\tiny {\textup{WF}}}}^{\textrm{ref}})$, computed over the full sky, as a function of scale, $\ell$, depicted in Figures \ref{fig:conv_test_M}, \ref{fig:conv_test_DM} and \ref{fig:conv_test_PCG} for the messenger, dual messenger and PCG algorithms, respectively. The relative errors on the small scales are of the order $10^{-12}$, $10^{-12}$ and $10^{-15}$, and conversely, on the large scales, $10^{-3}$, $10^{-3}$ and $10^{-2}$, correspondingly, for the messenger, dual messenger and PCG methods. The messenger technique displays smooth and nearly uniform convergence on small and intermediate scales, with the relative error dropping below $10^{-6}$, while remaining below $10^{-3}$ for the largest scales. For the dual messenger scheme, the corresponding convergence rate highlights the hierarchical fashion in which the solution is computed, while yielding similar final relative error across all scales as the messenger algorithm. The PCG method has the lowest relative error on the smallest scales, although this may be biased by the fact that the reference method is also a PCG, but remains inferior to both messenger methods on the largest scales. This is consistent with the significant residuals resulting from the PCG reconstruction, as observed in the previous section (cf. Figure \ref{fig:residual_map}). We stress that all relative errors above are computed with respect to the reference PCG method with $\epsilon = 10^{-9}$.

% Alternative position to insert three figures for convergence tests so that to get them all on same page.

\begin{figure}
	\centering
		{\includegraphics[width=\hsize,clip=true]{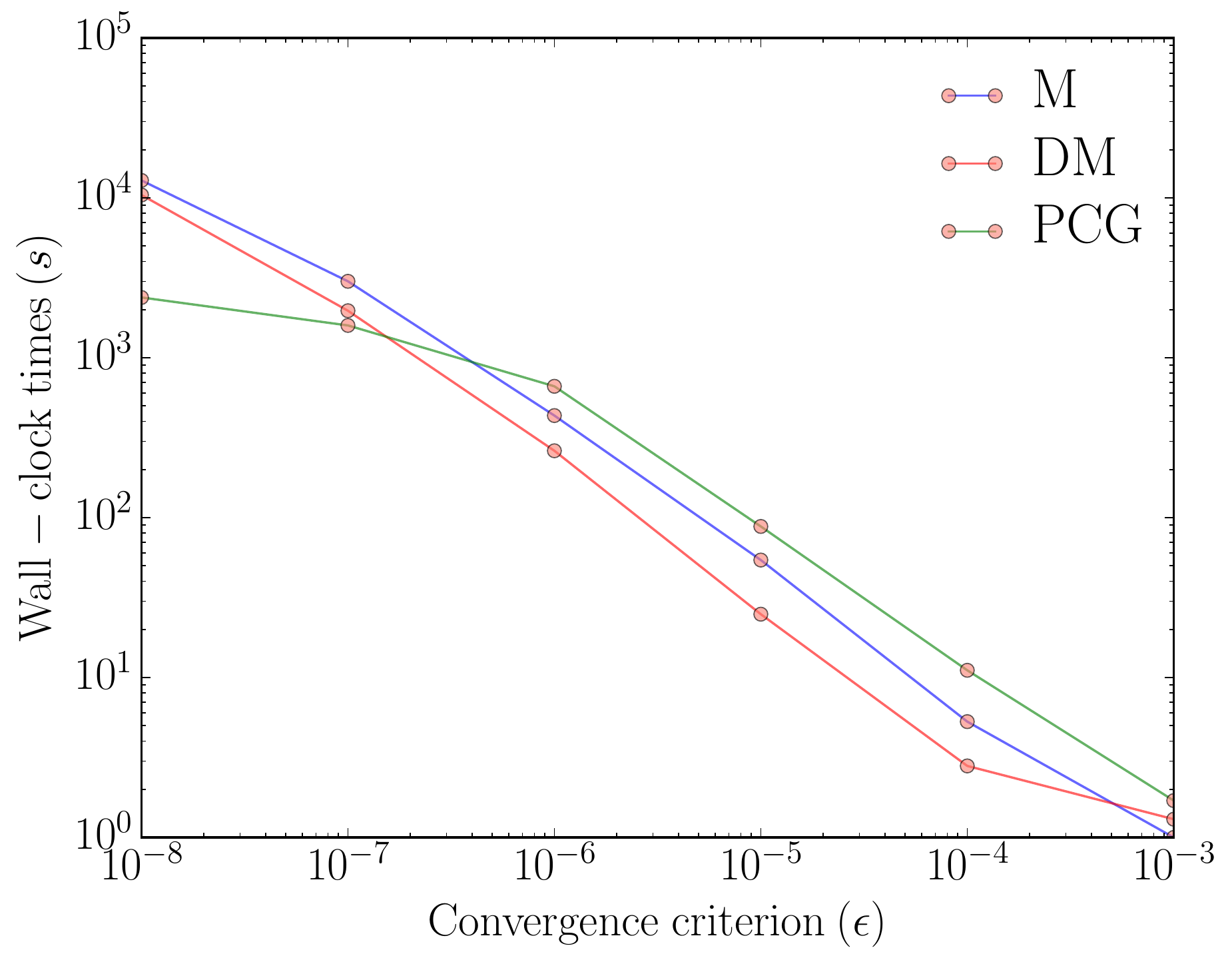}}
	\caption{Execution times required for convergence as a function of convergence criterion $\epsilon$ for the different methods. This demonstrates the efficient performance of the two messenger algorithms with respect to the PCG method.}
	\label{fig:plot_epsilon_times}
\end{figure}

We also carry out a series of additional runs with various values of the convergence criterion $\epsilon$ to  investigate the wall-clock times required for convergence in each case, thereby providing a more in-depth picture of the performance of the different algorithms. As illustrated in Figure \ref{fig:plot_epsilon_times}, the dual messenger technique converges faster than the other methods, except for the extreme values of $\epsilon$. It is also interesting to note that we can further reduce its execution time by lowering the factor $\beta$ for the cooling scheme without degrading the accuracy of results significantly. For instance, choosing $\beta = 1/2$ reduces the number of iterations required, and therefore computation time, by around $25 \%$ for the case $\epsilon = 10^{-6}$. 

\section{Conclusions}
\label{section5}

We presented a new formulation that is dual to the recently developed messenger algorithm, where, unlike in the standard approach, the auxiliary field is introduced at the level of the signal and is consequently associated with the signal reconstruction instead of the noise. This new iterative solver provides another pathway to solve the ill-posed problem of Wiener filtering, frequently encountered in several applications in cosmology and astrophysics. 

We tested our new method on a simulated CMB data set and its performance was evaluated in terms of effectiveness of reconstruction, convergence properties, processing time and stability. The dual messenger scheme is shown to match the accuracy of reconstruction of the standard messenger and PCG methods on all scales. Regarding the convergence of the algorithm, it is shown to perform smoothly over all scales, with the relative error being sufficiently low even on the largest scales. For the specific problem under consideration, the dual messenger algorithm has a definite edge over the standard messenger scheme and the PCG approach in terms of computation time. The efficiency and effectiveness of this new technique in calculating the Wiener filter solution of general data sets has therefore been demonstrated. We also provided further insight into the mechanism of the standard messenger method so that it can be optimised for data analysis by the scientific community.

The dual messenger algorithm, like its predecessor, does not require an ingenious choice of preconditioner and is straightforward to implement, robust and flexible. It is also capable of taking into account inhomogeneous noise distributions and arbitrary mask geometries. A key aspect of this technique is that it computes the Wiener filter solution in a hierarchical manner due to the thresholding of the signal covariance matrix inherent in the algorithm. Given the success of the standard messenger method, this new dual messenger technique is highly promising and may therefore be adapted for high precision large-scale data analysis methods in cosmology and astrophysics. 

Other performance improvements to this algorithm can be obtained by adapting the working resolution such that the Nyquist frequency is always slightly higher than the current $\ell_{\mathrm{iter}}$ considered in the dual messenger method. This would consequently reduce the number of operations required for Fourier transform. We are also currently working on a generalisation of both the messenger and dual messenger algorithms in a combined approach that may further optimise the performance and efficiency compared to traditional techniques of solving the Wiener filter problem. 

\section*{Acknowledgements}

The authors thank the anonymous reviewer for his constructive comments which helped to improve the manuscript. The authors also thank Franz Elsner for his comments on a draft version of the paper. The authors acknowledge financial support from the ILP LABEX (under reference ANR-10-LABX-63) which is financed by French state funds managed by the ANR within the Investissements d'Avenir programme under reference ANR-11-IDEX-0004-02. GL also acknowledges financial support from ``Programme National de Cosmologie et Galaxies'' (PNCG) of CNRS/INSU, France, and INPHYNITI programme of Interdisciplinary division of CNRS. 

\bibliographystyle{mn2e_guilhem}
%\bibliography{/home/doogesh/Bibliography/compiled_references} % if your bibtex file is called example.bib
\bibliography{messenger}

\begin{thebibliography}{}

\bibitem[\protect\citeauthoryear{}{{Alsing}
  et~al.}{2016a}]{alsing2016hierarchical}
{Alsing} J.,  {Heavens} A.,  {Jaffe} A.~H.,  {Kiessling} A.,  {Wandelt} B.,
  {Hoffmann} T.,  2016a, \mnras, 455, 4452, {arXiv:1505.07840}

\bibitem[\protect\citeauthoryear{}{{Alsing}
  et~al.}{2016b}]{alsing2016cosmological}
{Alsing} J.,  {Heavens} A.~F.,    {Jaffe} A.~H.,  2016b, ArXiv e-prints,
  {arXiv:1607.00008}

\bibitem[\protect\citeauthoryear{}{{Anderes}
  et~al.}{2015}]{anderes2015bayesian}
{Anderes} E.,  {Wandelt} B.~D.,    {Lavaux} G.,  2015, \apj, 808, 152,
  {arXiv:1412.4079}

\bibitem[\protect\citeauthoryear{}{{Bunn} \& {Wandelt}}{2016}]{bunn2016pure}
{Bunn} E.~F.,  {Wandelt} B.,  2016, ArXiv e-prints, {arXiv:1610.03345}

\bibitem[\protect\citeauthoryear{}{{Elsner} \&
  {Wandelt}}{2012}]{elsner2012fast}
{Elsner} F.,  {Wandelt} B.~D.,  2012, Proc. Big Bang, Big Data, Big Computers,
  {arXiv:1211.0585}

\bibitem[\protect\citeauthoryear{}{{Elsner} \& {Wandelt}}{2013}]{EW12}
{Elsner} F.,  {Wandelt} B.~D.,  2013, \aap, 549, A111, {arXiv:1210.4931}

\bibitem[\protect\citeauthoryear{}{{Erdo{\v g}du}
  et~al.}{2006}]{erdogdu2006reconstructed}
{Erdo{\v g}du} P.,  {Lahav} O.,  {Huchra} J.~P.,    {Colless} M.,  et~al. 2006,
  \mnras, 373, 45, {arXiv:astro-ph/0610005}

\bibitem[\protect\citeauthoryear{}{{Erdo{\v g}du}
  et~al.}{2004}]{erdogdu20042df}
{Erdo{\v g}du} P.,  {Lahav} O.,  {Zaroubi} S.,    {Efstathiou} G.,  et~al.
  2004, \mnras, 352, 939, {arXiv:astro-ph/0312546}

\bibitem[\protect\citeauthoryear{}{{Eriksen} et~al.}{2008}]{eriksen2008joint}
{Eriksen} H.~K.,  {Dickinson} C.,  {Jewell} J.~B.,  {Banday} A.~J.,
  {G{\'o}rski} K.~M.,    {Lawrence} C.~R.,  2008, \apjl, 672, L87,
  {arXiv:0709.1037}

\bibitem[\protect\citeauthoryear{}{{Eriksen} et~al.}{2004}]{eriksen2004power}
{Eriksen} H.~K.,  {O'Dwyer} I.~J.,  {Jewell} J.~B.,    {Wandelt} B.~D.,  et~al.
  2004, \apjs, 155, 227, {arXiv:astro-ph/0407028}

\bibitem[\protect\citeauthoryear{Gauss}{Gauss}{1809}]{gauss1809theoria}
Gauss C.~F.,  1809, Theoria motus corporum coelestium in sectionibus conicis
  solem ambientium auctore Carolo Friderico Gauss.
sumtibus Frid. Perthes et IH Besser

\bibitem[\protect\citeauthoryear{}{Golub \& Van~Loan}{1996}]{golub1996matrix}
Golub G.~H.,  Van~Loan C.~F.,  1996, Johns Hopkins University, Press,
  Baltimore, MD, USA, pp 374--426

\bibitem[\protect\citeauthoryear{}{{Hirata} et~al.}{2004}]{hirata2004cross}
{Hirata} C.~M.,  {Padmanabhan} N.,  {Seljak} U.,  {Schlegel} D.,    {Brinkmann}
  J.,  2004, \prd, 70, 103501, {arXiv:astro-ph/0406004}

\bibitem[\protect\citeauthoryear{}{{Jasche} et~al.}{2010}]{jasche2010bayesian}
{Jasche} J.,  {Kitaura} F.~S.,  {Wandelt} B.~D.,    {En{\ss}lin} T.~A.,  2010,
  \mnras, 406, 60, {arXiv:0911.2493}

\bibitem[\protect\citeauthoryear{}{{Jasche} \&
  {Lavaux}}{2015}]{jasche2015matrix}
{Jasche} J.,  {Lavaux} G.,  2015, \mnras, 447, 1204, {arXiv:1402.1763}

\bibitem[\protect\citeauthoryear{}{Jaynes}{1957}]{jaynes1957information}
Jaynes E.~T.,  1957, Physical review, 106, 620

\bibitem[\protect\citeauthoryear{}{{Jewell}
  et~al.}{2004}]{jewell2004application}
{Jewell} J.,  {Levin} S.,    {Anderson} C.~H.,  2004, \apj, 609, 1,
  {arXiv:astro-ph/0209560}

\bibitem[\protect\citeauthoryear{}{Kalman}{1960}]{kalman1960new}
Kalman R.~E.,  1960, Journal of basic Engineering, 82, 35

\bibitem[\protect\citeauthoryear{}{{Kitaura} \&
  {En{\ss}lin}}{2008}]{kitaura2008bayesian}
{Kitaura} F.~S.,  {En{\ss}lin} T.~A.,  2008, \mnras, 389, 497,
  {arXiv:0705.0429}

\bibitem[\protect\citeauthoryear{}{{Kitaura} et~al.}{2009}]{kitaura2009cosmic}
{Kitaura} F.~S.,  {Jasche} J.,  {Li} C.,  {En{\ss}lin} T.~A.,  {Metcalf} R.~B.,
   {Wandelt} B.~D.,  {Lemson} G.,    {White} S.~D.~M.,  2009, \mnras, 400, 183,
  {arXiv:0906.3978}

\bibitem[\protect\citeauthoryear{}{{Komatsu}
  et~al.}{2005}]{komatsu2005measuring}
{Komatsu} E.,  {Spergel} D.~N.,    {Wandelt} B.~D.,  2005, \apj, 634, 14,
  {arXiv:astro-ph/0305189}

\bibitem[\protect\citeauthoryear{}{{Larson}
  et~al.}{2007}]{larson2007estimation}
{Larson} D.~L.,  {Eriksen} H.~K.,  {Wandelt} B.~D.,  {G{\'o}rski} K.~M.,
  {Huey} G.,  {Jewell} J.~B.,    {O'Dwyer} I.~J.,  2007, \apj, 656, 653,
  {arXiv:astro-ph/0608007}

\bibitem[\protect\citeauthoryear{}{Lewis et~al.}{2000}]{camb1999}
Lewis A.,  Challinor A.,    Lasenby A.,  2000, Astrophys. J., 538, 473,
  {arXiv:astro-ph/9911177}

\bibitem[\protect\citeauthoryear{}{{Mangilli} \&
  {Verde}}{2009}]{mangilli2009nongaussianity}
{Mangilli} A.,  {Verde} L.,  2009, \prd, 80, 123007, {arXiv:0906.2317}

\bibitem[\protect\citeauthoryear{}{{Mangilli}
  et~al.}{2013}]{mangilli2013optimal}
{Mangilli} A.,  {Wandelt} B.,  {Elsner} F.,    {Liguori} M.,  2013, \aap, 555,
  A82, {arXiv:1303.1722}

\bibitem[\protect\citeauthoryear{}{{O'Dwyer} et~al.}{2004}]{odwyer2004bayesian}
{O'Dwyer} I.~J.,  {Eriksen} H.~K.,  {Wandelt} B.~D.,    {Jewell} J.~B.,  et~al.
  2004, \apjl, 617, L99, {arXiv:astro-ph/0407027}

\bibitem[\protect\citeauthoryear{}{{Oh} et~al.}{1999}]{oh1999efficient}
{Oh} S.~P.,  {Spergel} D.~N.,    {Hinshaw} G.,  1999, \apj, 510, 551,
  {arXiv:astro-ph/9805339}

\bibitem[\protect\citeauthoryear{}{{Planck Collaboration}
  et~al.}{2014a}]{23planck2013}
{Planck Collaboration} {Ade} P.~A.~R.,  {Aghanim} N.,    {Armitage-Caplan} C.,
  et~al. 2014a, \aap, 571, A23, {arXiv:1303.5083}

\bibitem[\protect\citeauthoryear{}{{Planck Collaboration}
  et~al.}{2014b}]{24planck2013}
{Planck Collaboration} {Ade} P.~A.~R.,  {Aghanim} N.,    {Armitage-Caplan} C.,
  et~al. 2014b, \aap, 571, A24, {arXiv:1303.5084}

\bibitem[\protect\citeauthoryear{}{{Planck Collaboration}
  et~al.}{2016}]{13planck2015}
{Planck Collaboration} {Ade} P.~A.~R.,  {Aghanim} N.,    {Arnaud} M.,  et~al.
  2016, \aap, 594, A13, {arXiv:1502.01589}

\bibitem[\protect\citeauthoryear{}{{Seljebotn}
  et~al.}{2014}]{seljebotn2014multi}
{Seljebotn} D.~S.,  {Mardal} K.-A.,  {Jewell} J.~B.,  {Eriksen} H.~K.,
  {Bull} P.,  2014, \apjs, 210, 24, {arXiv:1308.5299}

\bibitem[\protect\citeauthoryear{Shewchuk}{Shewchuk}{1994}]{shewchuk1994introduction}
Shewchuk J.~R.,  1994, An introduction to the conjugate gradient method without
  the agonizing pain.
Carnegie-Mellon University. Department of Computer Science

\bibitem[\protect\citeauthoryear{}{{Smith} et~al.}{2007}]{smith2007background}
{Smith} K.~M.,  {Zahn} O.,    {Dor{\'e}} O.,  2007, \prd, 76, 043510,
  {arXiv:0705.3980}

\bibitem[\protect\citeauthoryear{}{{Wandelt} et~al.}{2004}]{wandelt2004global}
{Wandelt} B.~D.,  {Larson} D.~L.,    {Lakshminarayanan} A.,  2004, \prd, 70,
  083511, {arXiv:astro-ph/0310080}

\bibitem[\protect\citeauthoryear{{Wiener}}{{Wiener}}{1949}]{wiener1949extrapolation}
{Wiener} N.,  1949, Extrapolation, interpolation, and smoothing of stationary
  time series.
Vol.~2, MIT press Cambridge

\bibitem[\protect\citeauthoryear{}{{Zaroubi}}{2002}]{zaroubi2002unbiased}
{Zaroubi} S.,  2002, \mnras, 331, 901, {arXiv:astro-ph/0010561}

\bibitem[\protect\citeauthoryear{}{{Zaroubi} et~al.}{1999}]{zaroubi1999wiener}
{Zaroubi} S.,  {Hoffman} Y.,    {Dekel} A.,  1999, \apj, 520, 413,
  {arXiv:astro-ph/9810279}

\end{thebibliography}

\appendix
\onecolumn 

\section{Cooling Scheme for Messenger Algorithm}
\label{cooling_scheme_appendix}

In this section, we illustrate the rationale behind the cooling scheme for the scalar parameter $\lambda$ adopted for the standard messenger technique. In the computations below, we assume invertibility of matrices everywhere. Using $\mymat{T} = \alpha \mathbb{1}$, where $\alpha = \textup{min}(\textup{diag}(\mymat{N}))$, the ${\chi^2_{\tiny {T}}}$ for the messenger scheme from Equation (\ref{eq:chi2_messenger}) can be written as
\begin{equation}
	{\chi^2_{\tiny {T}}} = (\myvec{d} - \myvec{s})^{\dagger} \left[ \mymat{N} + (\lambda - 1)\alpha\mathbb{1}\right]^{-1} (\myvec{d} - \myvec{s}) + \myvec{s}^{\dagger}\mymat{S}^{-1}\myvec{s}.
	\label{eq:chi2_appendix}
\end{equation}
Hence, in the messenger approach, we have $\mymat{N} \rightarrow [\mymat{N} + (\lambda - 1)\alpha \mathbb{1}]$ and plugging this into the Wiener filter given by Equation (\ref{eq:wf_equation}) yields
\begin{align*}
	\myvec{s}_{\text{\tiny {\textup{WF}}}}(\lambda) &=  \left\lbrace [ \mymat{N} + (\lambda - 1)\alpha\mathbb{1}]^{-1} + \mymat{S}^{-1} \right\rbrace^{-1} [\mymat{N} + (\lambda - 1)\alpha\mathbb{1}]^{-1} \myvec{d}\\ &= \mymat{S} \left[ \mymat{S} + \mymat{N} + (\lambda - 1)\alpha\mathbb{1} \right]^{-1} \myvec{d}, \numberthis
	\label{eq:WF_M_appendix}
\end{align*}
where we assume that all matrices are invertible. As a consistency check, for $\lambda = 1$, Equation (\ref{eq:WF_M_appendix}) reduces to Equation (\ref{eq:wf_equation}), the usual Wiener filter equation. The difference between the solutions at two consecutive values of $\lambda$ is given by
\begin{align*}
	\myvec{s}_{\text{\tiny {\textup{WF}}}}(\lambda) - \myvec{s}_{\text{\tiny {\textup{WF}}}}(\lambda') &= \mymat{S} \left\lbrace [\mymat{S} + \mymat{N} + (\lambda - 1)\alpha\mathbb{1}]^{-1} - [ \mymat{S} + \mymat{N} + (\lambda' - 1)\alpha\mathbb{1}]^{-1} \right\rbrace \myvec{d} \\ &= \mymat{S} [\mymat{S} + \mymat{N} + (\lambda - 1)\alpha\mathbb{1}]^{-1} \left\lbrace \mymat{S} + \mymat{N} + (\lambda' - 1)\alpha\mathbb{1} - [\mymat{S} + \mymat{N} + (\lambda - 1)\alpha\mathbb{1}]\right\rbrace [\mymat{S} + \mymat{N} + (\lambda' - 1)\alpha\mathbb{1}]^{-1} \myvec{d}\\ &= \alpha (\lambda' - \lambda) \mymat{S} \left[ \mymat{S} + \mymat{N} + (\lambda - 1)\alpha\mathbb{1} \right]^{-1} \left[ \mymat{S} + \mymat{N} + (\lambda' - 1)\alpha\mathbb{1} \right]^{-1} \myvec{d}. \numberthis
	\label{eq:WF_M_diff_appendix}
\end{align*}
Now, we consider limiting cases of Equation (\ref{eq:WF_M_diff_appendix}). We assume a homogeneous noise distribution, i.e., $\mymat{N} = \alpha \mathbb{1}$. From Equation (\ref{eq:WF_M_appendix}), we have
\begin{equation}
	\left\| \myvec{s}_{\text{\tiny {\textup{WF}}}} (\lambda) \right\|_{\myvec{\ell}} = C_{\myvec{\ell}} (C_{\myvec{\ell}} + \lambda \alpha)^{-1} \|\myvec{d}\|_{\myvec{\ell}}.
	\label{eq:norm_WF_appendix}
\end{equation}
Using Equations (\ref{eq:WF_M_diff_appendix}) and (\ref{eq:norm_WF_appendix}), the relative error can then be computed as follows:
\begin{align*}
	\frac{\left \| \myvec{s}_{\text{\tiny {\textup{WF}}}} (\lambda) - \myvec{s}_{\text{\tiny {\textup{WF}}}} (\lambda') \right \|_{\myvec{\ell}}}{\left \| \myvec{s}_{\text{\tiny {\textup{WF}}}} (\lambda) \right \|_{\myvec{\ell}}} &\leq \left[ |\lambda' - \lambda| \alpha \left( \frac{C_{\myvec{\ell}}}{C_{\myvec{\ell}} + \lambda\alpha} \right) \left( \frac{1}{C_{\myvec{\ell}} + \lambda' \alpha} \right) \|\myvec{d}\|_{\myvec{\ell}} \right] \left[\frac{C_{\myvec{\ell}}}{(C_{\myvec{\ell}} + \lambda \alpha)} \|\myvec{d}\|_{\myvec{\ell}} \right]^{-1} \\ & \leq \frac{\alpha|\lambda' - \lambda|}{C_{\myvec{\ell}} + \lambda'\alpha}, \numberthis
	\label{eq:relative_error_M_appendix}
\end{align*}
where, for an arbitrary matrix $\mymat{M}$, $\left \| \mymat{M} \right \|_{\myvec{\ell}}$ is the subset of $\left \| \mymat{M} \right \|$ over subspace $\myvec{\ell}$. From Equation (\ref{eq:residual_messenger1}), we have the fractional error reduction, after one iteration, given by
\begin{equation}
	1 -\epsilon \leq \frac{C_{\myvec{\ell}}}{C_{\myvec{\ell}} + \lambda \alpha} \left( \frac{\bar{\mymat{N}}}{\bar{\mymat{N}} + \lambda \alpha} \right),
	\label{eq:residual_M_appendix}
\end{equation}
where the term in parentheses will always favour rapid convergence. The other term is strongly convergent on small scales, but to give bounds to the maximum convergence speed of ``slow modes'', we focus on the latter, leading to 
\begin{equation}
	\epsilon \leq 1 - \frac{C_{\myvec{\ell}}}{C_{\myvec{\ell}} + \lambda \alpha} = \frac{\lambda \alpha}{C_{\myvec{\ell}} + \lambda \alpha}.
	\label{eq:epsilon_M_appendix}
\end{equation}
To favour convergence, from Equations (\ref{eq:relative_error_M_appendix}) and (\ref{eq:epsilon_M_appendix}), we require the two conditions:
\begin{equation}
	\frac{\alpha|\lambda' - \lambda|}{C_{\myvec{\ell}} + \lambda'\alpha} \la \left\{\begin{matrix}
\frac{\lambda \alpha}{C_{\myvec{\ell}} + \lambda \alpha} \\  \\
\frac{\lambda' \alpha}{C_{\myvec{\ell}} + \lambda' \alpha} \end{matrix}\right. 
	\label{eq:delta_lambda_M_appendix}
\end{equation}
so that the error reduction in one iteration matches the change in solution arising from two consecutive values of $\lambda$. This results in the following two constraints: $|\lambda'- \lambda| \leq \lambda'$ and $|\lambda'- \lambda| \leq \lambda$. The former is automatically satisfied, while the latter leads to the following bounds: $\lambda'/2 \leq \lambda \leq \lambda'$. Therefore, for the cooling scheme, we can write $\lambda = \eta \lambda'$, where $1/2 \leq \eta \leq 1$, to improve convergence on all scales. This is the motivation behind our choice of $\eta = 3/4$ in this work.

\section{Cooling Scheme for Dual messenger Algorithm}
\label{truncating_scheme_appendix}

As in the previous section, we wish to analytically determine the cooling scheme that would improve convergence for the dual messenger algorithm. We first derive the Wiener filter solution for the dual messenger, i.e., the analogue of Equation (\ref{eq:WF_M_appendix}) for this scheme. We begin by writing down Equations (\ref{eq:dual_messenger_1st_equation}) and (\ref{eq:dual_messenger_2nd_equation}), showing the dependence on the power spectrum truncation $\mu$:
\begin{align}
	\left(\mymat{N}^{-1} + \mymat{U}^{-1}_\mu \right)\myvec{s}_{\text{\tiny {\textup{WF}}}}(\mu) &= \mymat{N}^{-1}\myvec{d} + \mymat{U}^{-1}_\mu \myvec{u}_\mu \label{eq:dual_messenger_1st_equation_appendix}\\
	\left(\mymat{U}^{-1}_\mu + \bar{\mymat{S}}^{-1}_\mu \right)\myvec{u}_\mu &= \mymat{U}^{-1}_\mu \myvec{s}_{\text{\tiny {\textup{WF}}}}(\mu).
	\label{eq:dual_messenger_2nd_equation_appendix}
\end{align}
Solving for $\myvec{s}_{\text{\tiny {\textup{WF}}}}(\mu)$ via the following steps:
\begin{align*}
	\mymat{N}^{-1} \myvec{d} &= \left[ (\mymat{N}^{-1} + \mymat{U}^{-1}_\mu) - \mymat{U}^{-1}_\mu (\mymat{U}^{-1}_\mu + \bar{\mymat{S}}^{-1}_\mu)^{-1}\mymat{U}^{-1}_\mu \right] \myvec{s}_{\text{\tiny {\textup{WF}}}}(\mu)  \\  &= \left[(\mymat{N}^{-1} + \mymat{U}^{-1}_\mu) (\bar{\mymat{S}}_\mu  + \mymat{U}_\mu) - \bar{\mymat{S}}_\mu \mymat{U}^{-1}_\mu \right] (\bar{\mymat{S}}_\mu + \mymat{U}_\mu)^{-1} \myvec{s}_{\text{\tiny {\textup{WF}}}}(\mu) \\  &= \left[ \mymat{N}^{-1} (\bar{\mymat{S}}_\mu + \mymat{U}_\mu) + \mathbb{1} \right] (\bar{\mymat{S}}_\mu + \mymat{U}_\mu)^{-1} \myvec{s}_{\text{\tiny {\textup{WF}}}}(\mu) \\ &= \left[ \mymat{N}^{-1} + (\bar{\mymat{S}}_\mu + \mymat{U}_\mu)^{-1} \right]  \myvec{s}_{\text{\tiny {\textup{WF}}}}(\mu) , \numberthis
	\label{eq:workings_WF_DM_appendix}
\end{align*}
so that finally, we have
\begin{equation}
	\myvec{s}_{\text{\tiny {\textup{WF}}}}(\mu) = \left[ \mymat{N}^{-1} + (\bar{\mymat{S}}_\mu + \mymat{U}_\mu)^{-1} \right] \mymat{N}^{-1} \myvec{d}.
	\label{eq:WF_DM_appendix}
\end{equation}
In the limit $\mu \rightarrow \nu$, where $\nu \equiv \mathrm{min}(\textup{diag}(\mymat{S}))$, $\bar{\mymat{S}} + \mymat{U} \rightarrow \mymat{S}$, reducing Equation (\ref{eq:WF_DM_appendix}) to the usual Wiener filter Equation (\ref{eq:wf_equation}), implying consistency. When we change the truncation of the spectrum, $\mu \rightarrow \tilde{\mu}$, for $\tilde{\mu} < \mu$, we obtain a new solution $\myvec{s}_{\text{\tiny {\textup{WF}}}}(\tilde{\mu})$. The counterpart of Equation (\ref{eq:relative_error_M_appendix}) is then
\begin{align*}
	\myvec{s}_{\text{\tiny {\textup{WF}}}}(\tilde{\mu}) - \myvec{s}_{\text{\tiny {\textup{WF}}}}(\mu) &= (\bar{\mymat{S}}_{\tilde{\mu}} + \mymat{U}_{\tilde{\mu}})\left[ (\bar{\mymat{S}}_{\tilde{\mu}} + \mymat{U}_{\tilde{\mu}}) + \mymat{N} \right]^{-1} \myvec{d} - (\bar{\mymat{S}}_\mu + \mymat{U}_\mu)\left[ (\bar{\mymat{S}}_\mu + \mymat{U}_\mu) + \mymat{N} \right]^{-1} \myvec{d} \\ &= \left[ (\bar{\mymat{S}}_{\tilde{\mu}} + \mymat{U}_{\tilde{\mu}}) (\mymat{N} + \bar{\mymat{S}}_{\tilde{\mu}} + \mymat{U}_{\tilde{\mu}})^{-1} - (\bar{\mymat{S}}_\mu + \mymat{U}_\mu) (\mymat{N} + \bar{\mymat{S}}_\mu + \mymat{U}_\mu)^{-1} \right] \myvec{d}. \numberthis
	\label{eq:WF_DM_diff_appendix}
\end{align*}
Here, we consider Fourier space, so that $\mymat{U}_{\tilde{\mu}} \rightarrow \tilde{\mu}$ and $\mymat{U}_\mu \rightarrow \mu$. We can write the truncated signal covariance matrix as 
\begin{equation}
	\bar{\mymat{S}}_{\tilde{\mu}} = \bar{\mymat{S}}_{\mu} + \Delta_{\tilde{\mu},\mu}\mathbb{1},
	\label{eq:truncated_covariance_matrix}
\end{equation}
where $\Delta_{\tilde{\mu},\mu} = \tilde{\mu} - \mu$ is the portion of the power spectrum bounded by $\tilde{\mu}$ and $\mu$, while the corresponding truncated signal covariances can be represented by Heaviside functions as follows: 
\begin{align}
	\bar{\mymat{S}}_{\mu} &= \Theta(\mymat{S} - \mu) \label{eq:truncated_S1_appendix}\\
	\bar{\mymat{S}}_{\tilde{\mu}} &= \Theta(\mymat{S} - \tilde{\mu}),
	\label{eq:truncated_S2_appendix}
\end{align}
where, for a matrix $\mymat{M} = \mymat{P} \Lambda \mymat{P}^{-1}$, after applying a basis transformation, with $\Lambda$ being diagonal,
\begin{equation}
	\Theta(\mymat{M}) = \mymat{P}\Theta(\Lambda)\mymat{P}^{-1},
	\label{eq:heaviside1_appendix}
\end{equation}
and
\begin{equation}
	\Theta(\Lambda)_{ii} =   \left\{\begin{matrix}
0, \: \: \: \: \: \: \: \Lambda_{ii} \leq 0 \\ \Lambda_{ii}, \: \: \: \:\Lambda_{ii} > 0.
\end{matrix}\right.
	\label{eq:heaviside2_appendix}
\end{equation}
Using Equation (\ref{eq:truncated_covariance_matrix}) in Equation (\ref{eq:WF_DM_diff_appendix}) results in
\begin{align*}
	\myvec{s}_{\text{\tiny {\textup{WF}}}}(\tilde{\mu}) - \myvec{s}_{\text{\tiny {\textup{WF}}}}(\mu) &= \left[ (\bar{\mymat{S}}_{\mu} + \Delta_{\tilde{\mu},\mu} + \tilde{\mu}) (\mymat{N} + \bar{\mymat{S}}_{\mu} + \Delta_{\tilde{\mu},\mu} + \tilde{\mu})^{-1} - (\bar{\mymat{S}}_{\mu} - \mu) (\mymat{N} + \bar{\mymat{S}}_{\mu} + \mu)^{-1} \right] \myvec{d} \\ &= \left[ \bar{\mymat{S}}_{\mu} + \Delta_{\tilde{\mu},\mu} + \tilde{\mu} - (\bar{\mymat{S}}_\mu + \mu) (\mymat{N} + \bar{\mymat{S}}_\mu + \mu)^{-1} (\mymat{N} + \bar{\mymat{S}}_{\mu} + \Delta_{\tilde{\mu},\mu} + \tilde{\mu}) \right] (\mymat{N} + \bar{\mymat{S}}_{\mu} + \mu + \Delta_{\tilde{\mu},\mu} + \tilde{\mu} -\mu)^{-1} \myvec{d} \\ &= \left[ \bar{\mymat{S}}_{\mu} + \Delta_{\tilde{\mu},\mu} + \tilde{\mu} - \bar{\mymat{S}}_\mu - \mu - (\bar{\mymat{S}}_\mu + \mu) + (\mymat{N} + \bar{\mymat{S}}_\mu + \mu)^{-1} (\Delta_{\tilde{\mu},\mu} + \tilde{\mu} -\mu) \right] (\mymat{N} + \bar{\mymat{S}}_{\mu} + \Delta_{\tilde{\mu},\mu} + \tilde{\mu})^{-1} \myvec{d} \\ &= \left[ (\mymat{N} + \bar{\mymat{S}}_\mu + \mu) - \bar{\mymat{S}}_\mu - \mu \right] (\mymat{N} + \bar{\mymat{S}}_\mu + \mu)^{-1}(\Delta_{\tilde{\mu},\mu} + \tilde{\mu} -\mu) (\mymat{N} + \bar{\mymat{S}}_{\mu} + \Delta_{\tilde{\mu},\mu} + \tilde{\mu})^{-1} \myvec{d} \\ &= \mymat{N} (\mymat{N} + \bar{\mymat{S}}_\mu + \mu)^{-1}(\Delta_{\tilde{\mu},\mu} + \tilde{\mu} -\mu) (\mymat{N} + \bar{\mymat{S}}_{\mu} + \Delta_{\tilde{\mu},\mu} + \tilde{\mu})^{-1} \myvec{d} . \numberthis
	\label{eq:workings_WF_DM_diff_appendix}
\end{align*}
Using Equation (\ref{eq:truncated_covariance_matrix}) and substituting the following form of Equation (\ref{eq:WF_DM_appendix}),  
\begin{equation}
	\myvec{s}_{\text{\tiny {\textup{WF}}}}(\tilde{\mu}) = \bar{\mymat{S}}_{\tilde{\mu}} \left[ \mymat{N} + (\bar{\mymat{S}}_{\tilde{\mu}} + \tilde{\mu})^{-1} \right]^{-1} \myvec{d},
	\label{eq:rewritten_WF_DM_appendix}
\end{equation}
in Equation (\ref{eq:workings_WF_DM_diff_appendix}) leads to
\begin{equation}
	\myvec{s}_{\text{\tiny {\textup{WF}}}}(\tilde{\mu}) - \myvec{s}_{\text{\tiny {\textup{WF}}}}(\mu) = \mymat{N} (\mymat{N} + \bar{\mymat{S}}_\mu + \mu)^{-1}(\Delta_{\tilde{\mu},\mu} + \tilde{\mu} -\mu) (\bar{\mymat{S}}_{\tilde{\mu}} + \tilde{\mu})^{-1} \myvec{s}_{\text{\tiny {\textup{WF}}}}(\tilde{\mu}),
	\label{eq:final_workings_relative_error_DM_appendix}
\end{equation}
such that finally we obtain the following equation for the relative error: 
\begin{equation}
	\frac{\left \| \myvec{s}_{\text{\tiny {\textup{WF}}}}(\tilde{\mu}) - \myvec{s}_{\text{\tiny {\textup{WF}}}}(\mu) \right \|}{\left \| \myvec{s}_{\text{\tiny {\textup{WF}}}}(\tilde{\mu}) \right \|} \leq \left \| \mymat{N} (\mymat{N} + \bar{\mymat{S}}_\mu + \mu)^{-1}(\Delta_{\tilde{\mu},\mu} + \tilde{\mu} -\mu) (\bar{\mymat{S}}_{\tilde{\mu}} + \tilde{\mu})^{-1} \right \|.
	\label{eq:final_relative_error_DM_appendix}
\end{equation}
Now, for $\mu \rightarrow \tilde{\mu}$, there are three distinct regimes of relevance and we consider each case below. To investigate the convergence behaviour in the different regimes, we make use of the following equation, obtained by plugging Equations (\ref{eq:truncated_S1_appendix}) and (\ref{eq:truncated_S2_appendix}) in Equation (\ref{eq:truncated_covariance_matrix}),
\begin{equation}
	\Theta(\mymat{S} - \tilde{\mu}) = \Theta(\mymat{S} - \mu) + \Delta_{\tilde{\mu},\mu}.
	\label{eq:truncated_covariance_matrix2_appendix}
\end{equation}
The three regimes are:
\begin{itemize}[leftmargin=+.5in]
\item $\mymat{S}_{\myvec{\ell}} > \mu\mathbb{1}$, \\ where $\Theta(\mymat{S} - \tilde{\mu}) = \mymat{S}_{\myvec{\ell}} - \tilde{\mu}\mathbb{1}$, and $\Theta(\mymat{S} - \mu) = \mymat{S}_{\myvec{\ell}} - \mu\mathbb{1}$, so that $\Delta_{\tilde{\mu},\mu} = \mu - \tilde{\mu}$, and this causes the second term in Equation (\ref{eq:final_relative_error_DM_appendix}) to vanish, implying that the relative error is not affected by our choice of $\tilde{\mu}$. This shows that the Wiener filter solution is naturally computed in a hierarchical fashion using the dual messenger algorithm.
\item $\tilde{\mu}\mathbb{1} < \mymat{S}_{\myvec{\ell}} < \mu\mathbb{1}$,\\ where $\Theta(\mymat{S} - \tilde{\mu}) = \mymat{S}_{\myvec{\ell}} - \tilde{\mu}\mathbb{1}$, and $\Theta(\mymat{S} - \mu) = 0$, so that $\Delta_{\tilde{\mu},\mu} = \mymat{S}_{\myvec{\ell}} - \mu\mathbb{1}$, leading to the following relative error:
\begin{equation}
	\frac{\left \| \myvec{s}_{\text{\tiny {\textup{WF}}}}(\tilde{\mu}) - \myvec{s}_{\text{\tiny {\textup{WF}}}}(\mu) \right \|}{\left \| \myvec{s}_{\text{\tiny {\textup{WF}}}}(\tilde{\mu}) \right \|} \leq \left \| \mymat{N} (\mymat{N} + \bar{\mymat{S}}_\mu + \mu)^{-1}\right \| \left \|(\mymat{S}_{\myvec{\ell}} - \mu\mathbb{1}) (\bar{\mymat{S}}_{\tilde{\mu}} + \tilde{\mu})^{-1} \right \|_{\tilde{\mu}, \mu}.
	\label{eq:relative_error_regime2_appendix}
\end{equation}
The first term behaves as a constant, i.e., $\left \| \mymat{N} (\mymat{N} + \bar{\mymat{S}}_\mu + \mu)^{-1}\right \| \sim \alpha'$, so that the relative error can be approximated as
\begin{align*}
	\frac{\left \| \myvec{s}_{\text{\tiny {\textup{WF}}}}(\tilde{\mu}) - \myvec{s}_{\text{\tiny {\textup{WF}}}}(\mu) \right \|}{\left \| \myvec{s}_{\text{\tiny {\textup{WF}}}}(\tilde{\mu}) \right \|} &\leq \alpha' \left | \frac{C_{\myvec{\ell}} - \mu}{C_{\myvec{\ell}}} \right |_{\tilde{\mu}, \mu} \\ &=  \alpha' \left | 1 - \frac{\mu}{C_{\myvec{\ell}}} \right |_{\tilde{\mu}, \mu}, \numberthis
	\label{eq:approximated_relative_error_regime2_appendix}
\end{align*}
since $\bar{\mymat{S}}_{\tilde{\mu}} + \tilde{\mu} \sim C_{\myvec{\ell}}$. Also, $\alpha' \sim 1$, and to favour convergence, we want this change to be as large as possible but sufficiently small such that iterating the solution results in rapid decay of the error, i.e., the change in the solution due to changing $\mu$ should be matched to the change when iterating the solution. If $\mu = \beta C_{\myvec{\ell}}$, choosing $0 < \beta < 1$ would therefore improve convergence by avoiding the early freeze of modes in the iteration scheme. This served as the basis for our choice of $\beta = 3/4$ in this work.
\item $\mymat{S}_{\myvec{\ell}} < \tilde{\mu}\mathbb{1}$,\\ where $\Theta(\mymat{S} - \tilde{\mu}) = 0 = \Theta(\mymat{S} - \mu)$, so that $\Delta_{\tilde{\mu},\mu} = 0$, and again using the approximation $\bar{\mymat{S}}_{\tilde{\mu}} + \tilde{\mu} \sim C_{\myvec{\ell}}$ yields 
\begin{equation}
	\frac{\left \| \myvec{s}_{\text{\tiny {\textup{WF}}}}(\tilde{\mu}) - \myvec{s}_{\text{\tiny {\textup{WF}}}}(\mu) \right \|}{\left \| \myvec{s}_{\text{\tiny {\textup{WF}}}}(\tilde{\mu}) \right \|} \leq \alpha' \left| \frac{\tilde{\mu} - \mu}{C_{\myvec{\ell}}} \right|_{\tilde{\mu}, \mu}.
	\label{eq:approximated_relative_error_regime3_appendix}
\end{equation}
\end{itemize}
So, the overall convergence behaviour can be quantitatively described by:
\begin{equation}
	\frac{\left \| \myvec{s}_{\text{\tiny {\textup{WF}}}}(\tilde{\mu}) - \myvec{s}_{\text{\tiny {\textup{WF}}}}(\mu) \right \|}{\left \| \myvec{s}_{\text{\tiny {\textup{WF}}}}(\tilde{\mu}) \right \|} \leq 0 + \alpha' \left | 1 - \frac{\mu}{C_{\myvec{\ell}}} \right |_{\tilde{\mu}, \mu} + \alpha' \left| \frac{\tilde{\mu} - \mu}{C_{\myvec{\ell}}} \right|_{\tilde{\mu}, \mu},
	\label{eq:overall_relative_error_DM_appendix}
\end{equation}
and this leads to the following interpretation: For higher values of $\mu$ on large scales, the third term in Equation (\ref{eq:overall_relative_error_DM_appendix}) dominates since $\left| \tilde{\mu} - \mu \right|$ is large. But for the final truncations, $\tilde{\mu} \sim \mu$, so this regime is saturated, while on small scales, the second term dominates as $\mu \gg C_{\myvec{\ell}}$, so the relative error continues to drop till the final truncation. The cooling scheme described above applies naturally to the hybrid version of the dual messenger method.

\section{Brief Review of Preconditioned Conjugate Gradient Method}
\label{pcg_appendix}

In general, the PCG approach consists of solving the following set of linear equations:
\begin{equation}
	\mymat{\mathcal{A}}\myvec{x} = \mymat{y},
	\label{eq:pcg1}
\end{equation}
where $\mymat{\mathcal{A}}$ is usually a very large matrix. We wish to avoid computing the inverse of this dense matrix and to this end, in the traditional PCG scheme, we make use of a sparse matrix known as the preconditioner $\mymat{\mathcal{M}}$, which is the approximate inverse of $\mymat{\mathcal{A}}$, i.e., $\mymat{\mathcal{M}} \approx \mymat{\mathcal{A}}^{-1}$, as follows:
\begin{equation}
	\mymat{\mathcal{M}}\mymat{\mathcal{A}}\myvec{x} = \mymat{\mathcal{M}} \mymat{y}.
	\label{eq:pcg2}
\end{equation}
The Wiener filter Equation (\ref{eq:wf_equation}) can be rewritten as
\begin{equation}
	(1 + \mymat{S}^{\frac{1}{2}}\mymat{N}^{-1}\mymat{S}^{\frac{1}{2}})\mymat{S}^{-\frac{1}{2}}\myvec{s}_{\text{\tiny {\textup{WF}}}} = \mymat{S}^{\frac{1}{2}} \mymat{N}^{-1}\myvec{d},
	\label{eq:wf_equation_mod}
\end{equation}
such that $\mymat{\mathcal{A}} = 1 + \mymat{S}^{\frac{1}{2}}\mymat{N}^{-1}\mymat{S}^{\frac{1}{2}}$, $\myvec{x} = \mymat{S}^{-\frac{1}{2}}\myvec{s}_{\text{\tiny {\textup{WF}}}}$ and $\mymat{y} = \mymat{S}^{\frac{1}{2}} \mymat{N}^{-1}\myvec{d}$, in accordance with Equation (\ref{eq:pcg1}).
For the case considered in this work, we make use of a suitable preconditioner, which is diagonal in Fourier space, as follows:
\begin{equation}
	\mymat{\mathcal{M}}^{-1}_{\myvec{\ell}\myvec{\ell}} = \mymat{\mathcal{A}}_{\myvec{\ell}\myvec{\ell}} = \left( \sum_{\myvec{k}} n_{\myvec{k}} \right) C_{\myvec{\ell}} \frac{1}{L^4},
	\label{eq:preconditioner}
\end{equation}
where $n_{\myvec{k}}$ are the eigenvalues of the inverse noise covariance matrix, $\mymat{N}^{-1}$. We then implement this preconditioner in a PCG algorithm \citep*[e.g.,][]{golub1996matrix}. We make an initial guess $\myvec{x}_0$ and interate until $\left| \myvec{x}_{i+1} - \myvec{x}_{i} \right|/\left| \myvec{x}_{i} \right| < \epsilon$, resulting in an $\myvec{x}_{\text{\tiny {\textup{WF}}}}$ that corresponds to the Wiener filter solution $\myvec{s}_{\text{\tiny {\textup{WF}}}}$. If convergence is not yet achieved, we re-initialise all parameters every 200 iterations and resume iterations, in order to facilitate convergence. An in depth review of the PCG algorithm is provided in \cite{shewchuk1994introduction}. Again, we stress the fact that finding a suitable preconditioner is the key factor when making use of the PCG method \citep*[e.g.,][]{oh1999efficient} and this consequently is the major stumbling block for state-of-the-art CMB data analysis.

% Don't change these lines
\bsp	% typesetting comment
\label{lastpage}
\end{document}